\documentclass{aa}
\usepackage{graphicx}
\usepackage{txfonts}
\usepackage{lscape}
\usepackage{multirow}
\usepackage{xcolor}
\usepackage{ulem}
\usepackage{comment}
\usepackage{natbib}
\usepackage{amssymb}
\usepackage{scrextend} 
\usepackage{placeins} 
\usepackage{hyperref}
\hypersetup{
    colorlinks=true,
    linkcolor=blue,
    filecolor=magenta,
    urlcolor=blue,
    citecolor=blue,
} 

\begin{document} 

   \title{New stellar bow shocks and bubbles found around runaway stars}

   \author{M. Carretero-Castrillo
          \inst{1}
          \and P. Benaglia \inst{2}
          \and J. M. Paredes \inst{1}
          \and M. Rib\'o \inst{1}\fnmsep\thanks{Serra H\'unter Fellow}
          }

   \institute{Departament de Física Quàntica i Astrofísica, Institut de Ciències del Cosmos (ICCUB), Universitat de Barcelona, IEEC-UB, Martí i Franquès 1, 08028 Barcelona, Spain\\
   \email{mcarretero@fqa.ub.edu}
   \and Instituto Argentino de Radioastronomía (CONICET--CICPBA--UNLP), C.C No 5. 1894 Villa Elisa, Argentina\\
   \email{paula@iar-conicet.gov.ar}}

   \date{Received 1 July 2024 / Accepted 5 December 2024}

 
  \abstract
   {Runaway stars with peculiar high velocities can generate stellar bow shocks. Only a few bow shocks show clear radio emission.}
   {Our goal is to identify and characterize new stellar bow shocks around O and Be runaway stars in the infrared (IR), as well as to study their possible radio emission and nature.}
   {Our input data is a catalog of O and Be runaways compiled using \textit{Gaia}~DR3. We used WISE IR images to search for bow shock structures around these runaways, \textit{Gaia}~DR3 data to determine the actual motion of the runaway stars corrected for interstellar medium (ISM) motion caused by Galactic rotation, and archival radio data to search for emission signatures. We finally explored the radio detectability of these sources under thermal and nonthermal scenarios.}
   {We found nine new stellar bow shock candidates, three new bubble candidates, and one intermediate structure candidate. One of them is an in situ bow shock candidate. We also found 17 already known bow shocks in our sample, though we discarded one, and 62 miscellaneous sources showing some IR emission around the runaways. We geometrically characterized the sources in IR using the WISE-4 band and estimated the ISM density at the bow shock positions, obtaining median values of $\sim$6 and $\sim$4~cm$^{-3}$ using 2D and 3D peculiar velocities, respectively. Most of the new discovered bow shocks come from new runaway discoveries. Within our samples we found that $\sim$24\% of the O-type runaway stars show bow shocks, while this decreases to $\sim$3\% for Be-type runaway stars. Two bow shocks present radio emission but not as clear counterparts, and two others show hints of radio emission. The physical scenarios indicate that two sources could still be compatible with nonthermal radio emission.}
   {The new sample of O and Be runaway stars allowed us to discover both new stellar bow shocks and bubbles. Their geometrical characterization can be used to assess the physical scenario of the radio emission. Deeper radio observations are needed to unveil a population of radio-detected bow shocks, and the physical scenarios occurring in these sources.}


   \maketitle


\section{Introduction} \label{Sec:Introduction}

Runaway stars move with a high peculiar velocity relative to their environment (see \citet{Blaauw1961} for a seminal work, and \citet{MCC2023} for an updated reference). Aside from the studies of the origin of their runaway status, there is a whole area of research related to the phenomena that occur when a runaway star, through its strong stellar wind, interacts with the interstellar medium (ISM).

Stars of spectral types O and early B are among those with the most intense stellar winds. About 30\% of O stars and 5--10\% of B stars have been found to be runaway stars \citep{Blaauw1961,Stone1979,MCC2023}. Detailed studies to identify runaway stars have been performed by several authors \citep{Tetzlaff2011, Boubert2018, MA2018, Kobulnicky2022, MCC2023}. In this context, a stellar bow shock is formed by the accumulation of ISM matter as a runaway star with a powerful wind passes through the ISM \citep[first findings, e.g.][]{Gull1979}. If the motion is supersonic (i.e., the peculiar velocity of the star is greater than the speed of sound in the environment that is typically of some kilometers per second) strong shocks take place around a discontinuity surface separating the two media \citep[see for instance][and references therein]{Mackye2023}.

The swept-up matter, formed by warm dust and heated by stellar ultraviolet photons, shines at infrared (IR) wavelengths. The bow shocks usually show an arc-shaped structure but sometimes even a bubble surrounding the star, where part of the edge can be enhanced (that in the direction of the stellar motion). Initial searches for nearby runaway stars were carried out by \citet{NoriegaCrespo1997} (see also references therein). The authors studied high-resolution IRAS satellite images around runaway OB stars and published the first catalog of 21 possible bow shocks and 4 possible bubbles. One of them, later named EB27 and associated with the star \object{BD$+$43~3654}, was well resolved with IR data from the MSX satellite by \citet{Comeron2007}. More recently, four relevant catalogs of stellar bow shocks with IR emission have been compiled. The first one compiled by \citet{Peri2012,Peri2015}, called the Extensive stellar Bow Shock Survey (EBOSS), presented 73 object candidates in WISE images. Almost half of them had an associated runaway star. For the rest, there was either no information on the velocity of the star (i.e., its runaway status) that could be associated with the bow shock, or, in the case of a few of them, no associated star. The second catalog was compiled by \citet{Kobulnicky2016}, with 709 candidates detected using WISE and Spitzer images. In \citet{Kobulnicky2017}, the team publish the photometric data, and in \citet{Chick2020}, they study the 84 -- out of 709 -- cases in which stars have been identified supporting the bow-shock nebulae. The third one, \cite{Bodensteiner2018}, was compiled from different O, B, Be, and A-type star catalogs, but for general IR nebulae (including bow shocks) around massive stars. In this work, they test if the IR nebulae information could be a binary indicator, and confirm that for about 29\% of their sample. The fourth one, \cite{Jayasinghe2019}, contains 1394 bubbles and 453 bow shocks, compiled through the Milky Way Project using {\it Spitzer} data.

In terms of theoretical developments, the shape, dynamics, and emission of bow shocks have been modeled. \citet{Wilkin1996} presented one of the first solution methods that provided, among other parameters, the shape of the shell, the mass column density, and the velocity distribution of the shocked gas within it. The spectral energy distribution (SED) down to gamma rays of stellar bow shocks formed by O-type runaway stars was computed by \citet{Delvalle2012} using $\rho$~Oph as a test case. \citet{delpalacio2018} developed a multi-zone stellar bow shock model and SED for different scenarios, along with synthetic radio maps to reproduce the morphology of EB27 and predicted the high-energy (HE) emission. Using deep Very Large Array (VLA) low-frequency observations of the same target, \citet{Benaglia2021} model the intensity and morphology of the radio emission.

Strong bow shocks are promising places for the acceleration of particles up to relativistic velocities. Relativistic particles radiate across the electromagnetic spectrum, particularly in the radio, X-rays, and gamma rays \citep{delpalacio2018,Martinez2023}. In this context, \cite{Binder2019} searched for X-ray emission after stacking archival {\it Chandra} observations from 60 IR-bright Galactic bow shocks. Although there was no detection, the authors could establish a constraining upper limit on the X-ray luminosity of IR-detected bow shocks of $< 2 \times 10^{29}$~erg~s$^{-1}$, which is on the same order as model predictions. The question of whether particles can be accelerated up to relativistic energies at the shocks of these structures was investigated in the radio domain by \citet{Benaglia2010}. They performed L- and C-band radio observations of the source EB27 with the VLA, to look for nonthermal radio emission. Their ultimate goal was to determine whether bow shocks could be the counterparts of HE gamma-ray sources, especially of the unassociated ones. Radio emission coinciding with the IR extension of the EB27 bow shock was detected in both bands, and a central part of it showed a spectral index, $\alpha$ (with $S_{\nu} \sim \nu^{\alpha}$), close to $-0.5$, which is characteristic of nonthermal emission. Follow-up JVLA observations in the S-band in full polarization mode showed no polarized emission above the 1\% degree level \citep{Benaglia2021}. \cite{Martinez2023} have studied the nature of the radio emission for EB27 using a multi-zone model. Compared with radio observations, they conclude that the emission of EB27 can be explained through nonthermal electrons following a hard energy distribution below $\sim$1~GeV. Nevertheless, the emission from EB27 together with its spectral index and polarization absence needs further investigation.

In recent years, the commissioning of interferometers in the southern hemisphere has allowed several radio detections of stellar bow shocks and spectral index studies. \citet{VandenEijnden2022a} detected the Vela X-1 bow shock in radio with MeerKAT and found that its multiwavelength emission is more compatible with thermal free-free emission than with nonthermal synchrotron emission. \citet{VandenEijnden2022b} examined the Rapid ASKAP Continuum Survey \citep[RACS,][]{RACS2020} at the positions of 50 EBOSS candidates and detected 10 structures, classifying them as 3 radio bow shocks, 3 probable radio bow shocks, 1 to be confirmed, and 3 unclear (probably not radio bow shocks). The analysis of the first seven structures yielded results consistent with nonthermal emission for two of them. In addition, \citet{Moutzouri2022} measured a spectral index below $-0.5$ for the northern of the Bubble Nebula (NGC 7635, most likely formed by the runaway star BD$+$60 2522), which is the second bow shock with clear evidence for nonthermal emission. However, the question of bow shocks capable of accelerating relativistic particles and producing HE emission remains open.

The runaway stars associated with the bow shocks may also produce radio emission. In binary systems containing a massive O or Be star and a compact object, nonthermal radio emission can be produced by synchrotron-emitting relativistic electrons. There are two main types of these binary systems: a) X-ray binaries with a radio jet or microquasars \citep{Mirabel1999,Fender2016}; and b) gamma-ray binaries \citep{Dubus2006,Valenti2012, Dubus2013}. For the latter, three of the nine confirmed TeV-emitting gamma-ray binaries are runaways: \object{LS~5039}, \object{PSR~B1259$-$63} and \object{1FGL~J1018.6$-$5856} \citep{Ribo2002,Miller-Jones2018,Marcote2018}. Therefore, new gamma-ray binaries could be found by searching for massive runaway stars and their associated multiwavelength emission. This strategy could provide new members of this reduced sample, eventually allowing for the population studies that are needed to answer some of the many open questions in this area of HE astrophysics \citep[see][and references therein]{Bordas2023}.

For all of the above reasons, any study of stellar bow shocks and their radio counterparts can add information on a wide variety of phenomena. This is especially true if the runaway star that produces the bow shock is identified. In this work, we aim to detect bow shocks around new runaway stars and describe their characteristics. In Sect.~\ref{Sec:Data}, we present the data used in this work. In Sect.~\ref{Sec:Methodology}, we describe the methodology of the search, and the identification and characterization of the objects. In Sect.~\ref{Sec:Results}, we present the results for the different types of objects we found. We discuss these results in Sect.~\ref{Sec:Discussion}. We close with a summary and prospects for further studies in Sect.~\ref{Sec:Summary}.


\section{Data} \label{Sec:Data}

\subsection{Runaway star catalogs}

\renewcommand{\arraystretch}{1.1}
\begin{table*}
\centering
\caption{Data of the 10 out of 106 GOSC-\textit{Gaia}~DR3 runaway stars with higher values of $E$ in decreasing order.}
\label{Tab:GOSC_Runaways}
\resizebox{\textwidth}{!}{\begin{tabular}{l@{~~~}l@{~~~}c@{~~~}r@{~}c@{~}l@{~~~}r@{~}c@{~}l@{~~~}c@{~~~}c@{~~~}c@{~~~}c@{~~~}c@{~~~}r@{$~\pm~$}l@{~~~}c@{~~~}c@{~~~}c@{~~~}}
\hline\hline \vspace{-2mm}\\
Runaway id & GOSC Name & \textit{Gaia} DR3 id & \multicolumn{3}{c}{RA} & \multicolumn{3}{c}{DEC} & $d$ & $\mu_{\alpha,\text{corr}}^{*}$ & $\mu_{\delta,\text{corr}}$ & $G$ & S.T.& \multicolumn{2}{c}{$V_\text{PEC}^\text{2D}$} & $E$ & Run. Ref. 
 & Bow. Ref.\\
& & & \multicolumn{3}{c}{(h m s)} & \multicolumn{3}{c}{($\degr$  $\arcmin$  $\arcsec$)} & (kpc) &(mas yr$^{-1}$) & (mas yr$^{-1}$)& & & \multicolumn{2}{c}{(km~s$^{-1}$)} &\\
\hline \vspace{-2mm}\\
  GR1 & V479 Sct       & 4104196427943626624 & 18 & 26 & 15.06 & $-$14 & 50 & 54.4 & 1.94   & ~~~7.178    & $-$6.866 &  10.80 & ON6   & 90.9  & 4.9     & 4.28 & 1 &  ---\\
  GR2 & CPD $-$34 2135 & 5546501254035205376 & 08 & 13 & 35.36 & $-$34 & 28 & 43.9 & 3.23   & ~~~5.848    & $-$2.894 & ~~9.18 & O7.5  & 85.1  & 5.6     & 3.85 & 2 &  ---\\
  GR3 & BD $+$60 134   & 427457895747434880  & 00 & 56 & 14.21 &    61 & 45 & 36.9 & 2.69   & ~~~2.928    & $-$5.840 &  10.40 & O5.5  & 77.9  & 4.8     & 3.70 & 3 &  ---\\
  GR4 & HD 104 565     & 6072058878595295488 & 12 & 02 & 27.77 & $-$58 & 14 & 34.3 & 4.72   & $-$8.694    & ~~~4.853 & ~~9.07 & OC9.7 & 193.7 & 25.2    & 3.47 & 2 &  ---\\
  GR5 & HD 155 913     & 5953699131931631232 & 17 & 16 & 26.32 & $-$42 & 40 & 04.0 & 1.21   & $-$9.539    & ~~~7.676 & ~~8.18 & O4.5  & 70.6  & 4.5     & 3.43 & 2 &  ---\\
  GR6 & HD 75 222      & 5625488726258364544 & 08 & 47 & 25.13 & $-$36 & 45 & 02.5 & 2.09   & ~~~0.123    & ~~~6.320 & ~~7.31 & O9.7  & 67.4  & 6.9     & 3.03 & 2 &  ---\\
  GR7 & ALS 12 688     & 2003815312630659584 & 22 & 55 & 44.94 &    56 & 28 & 36.6 & 4.34   & ~~~4.061    & $-$1.815 &  10.39 & O5.5  & 74.5  & 6.8     & 2.90 & 3 &  ---\\
  GR8 & BD $-$14 5040  & 4104201586232296960 & 18 & 25 & 38.91 & $-$14 & 45 & 05.8 & 1.63   & ~~~5.548    & $-$3.423 &  10.02 & O5.5  & 50.4  & 2.3     & 2.87 & 2 &  I, II\\
  GR9 & HD 41 997      & 3345950879898371712 & 06 & 08 & 55.82 &    15 & 42 & 18.0 & 1.75   & $-$0.586    & $-$9.374 & ~~8.30 & O7.5  & 77.3  & 9.1     & 2.79 & 2 &  ---\\
  GR10 & Y Cyg         & 1869256701670871168 & 20 & 52 & 03.58 &    34 & 39 & 27.2 & 1.39   & ~~~4.731    & $-$11.691& ~~7.26 & O9.5  & 82.1  & 10.0    & 2.78 & 2 &  ---\\
\hline
\end{tabular}}
\tablefoot{Run. Ref. (Runaway Reference) column provides a reference in case the star had already been reported as runaway: (1)~\citet{Ribo2002}; (2) \citet{MA2018}; (3) \citet{Kobulnicky2022}. Bow. Ref. (Bowshock Reference) column provides a reference in case the star had already been reported as having a bow shock: (I)~\citet{Peri2015}; (II)~\citet{Kobulnicky2016}. The proper motions corrected for the ISM motion caused by Galactic rotation (see text) are $\mu_{\alpha,\text{corr}}^{*}$ and $\mu_{\delta,\text{corr}}$ (with $\mu_{\alpha,\text{corr}}^{*} = \mu_{\alpha,\text{corr}}\cos{\delta}$). The complete version of this table including the 106 classified runaways is available at the CDS.}
\end{table*}

\renewcommand{\arraystretch}{1.1}
\begin{table*}
\centering
\caption{Data of the 10 out of 69 BeSS-\textit{Gaia}~DR3 runaway stars with higher values of $E$ in decreasing order.} 
\label{Tab:BeSS_Runaways}
\resizebox{\textwidth}{!}{\begin{tabular}{l@{~~~}l@{~~~}c@{~~~}r@{~}c@{~}l@{~~~}r@{~}c@{~}l@{~~~}c@{~~~}c@{~~~}c@{~~~}c@{~~~}c@{~~~}r@{$~\pm~$}l@{~~~}c@{~~~}c@{~~~}c@{~~~}}
\hline\hline \vspace{-2mm}\\
Runaway id & BeSS Name & \textit{Gaia} DR3 id & \multicolumn{3}{c}{RA} & \multicolumn{3}{c}{DEC} & $d$ & $\mu_{\alpha,\text{corr}}^{*}$ & $\mu_{\delta,\text{corr}}$ & $G$ & S.T.& \multicolumn{2}{c}{$V_\text{PEC}^\text{2D}$} & $E$ & Run. Ref. 
 & Bow. Ref.\\
& & & \multicolumn{3}{c}{(h m s)} & \multicolumn{3}{c}{($\degr$  $\arcmin$  $\arcsec$)}  & (kpc) &(mas yr$^{-1}$) & (mas yr$^{-1}$)& & & \multicolumn{2}{c}{(km~s$^{-1}$)} &\\
\hline \vspace{-2mm}\\
  BR1 & CD$-$29 6963 & 5641493526747324416 & 08 & 56 & 48.24  & $-$30 & 11 & 01.8 & 3.70 & ~~~~4.128  & ~~$-$0.848  & ~~9.15  & Be  & 60.2 & 3.8   & 2.85 & --- &  ---\\
  BR2 & HD 181409    & 2049168552364523520 & 19 & 19 & 03.72  &    33 & 23 & 19.2 & 0.55 & $-$13.680  & $-$30.266  & ~~6.55  & B2e  & 85.3  & 5.8  & 2.64 & 1  &  ---\\
  BR3 & HD 114200    & 5843657293789559296 & 13 & 10 & 52.71  & $-$70 & 48 & 31.1 & 2.29 & ~~~~5.223  & ~~$-$2.846 & ~~8.35  & B1e  & 74.7  & 8.0  & 2.55 & --- &  ---\\
  BR4 & BD$-$21 1449 & 2938176961013250432 & 06 & 25 & 16.94  & $-$21 & 20 & 04.3 & 2.70 & ~~$-$3.596 & ~~~~0.354  &  10.43  & Be  & 52.3  & 5.7  & 2.44 & --- &  ---\\
  BR5 & EM* StHA 143 & 1333493685058234880 & 17 & 14 & 25.38  &    31 & 35 & 00.4 & 4.12 & ~~~~15.466 & $-$13.428  &  11.70  & Be  & 131.7 & 17.5 & 2.33 & --- &  ---\\
  BR6 & HD 77147     & 5297543543432635008 & 08 & 57 & 23.29  & $-$63 & 29 & 49.1 & 0.68 & ~~$-$1.481 & ~~$-$7.969 & ~~8.42  & B8  & 66.1  & 2.2  & 2.33 & --- &  ---\\
  BR7 & HD 127617    & 1239086696118147712 & 14 & 31 & 59.56  &    18 & 46 & 00.5 & 1.71 & ~~~~0.228 & $-$13.091  & ~~8.71  & B5e & 105.7 & 16.2 & 2.28 &  2 &  ---\\
  BR8 & BD$-$11 2043 & 3033914530822194688 & 07 & 39 & 42.24  & $-$12 & 15 & 33.3 & 3.64 & ~~~~1.111  & ~~~~1.962  & ~~9.43  & Be  & 52.7  & 9.6  & 2.16 & --- &  ---\\
  BR9 & HD 43789     & 5567419359659568896 & 06 & 15 & 56.55  & $-$44 & 37 & 10.6 & 0.92 & ~~$-$9.042 & ~~~~1.732  & ~~8.51  & B6.5 & 38.5 & 3.6  & 2.07 & --- &  ---\\
  BR10 & HD 159489    & 5955254318081779072 & 17 & 37 & 13.83 & $-$45 & 09 & 26.7 & 1.05 & ~~$-$6.572 & ~~~~1.668  & ~~8.15  & B1e  & 33.6  & 3.1  & 2.07 & --- &  ---\\
\hline
\end{tabular}}
\tablefoot{Run. Ref. (Runaway Reference) column provides a reference in case the star had already been reported as runaway: (1)~\citet{Hoogerwerf2001}; (2) \citet{Boubert2018}. Bow. Ref. (Bowshock Reference) column provides a reference in case the star had already been reported as having a bow shock. The proper motions corrected for the ISM motion caused by Galactic rotation (see text) are $\mu_{\alpha,\text{corr}}^{*}$ and $\mu_{\delta,\text{corr}}$ (with $\mu_{\alpha,\text{corr}}^{*} = \mu_{\alpha,\text{corr}}\cos{\delta}$). The complete version of this table including the 69 classified runaways is available at the CDS.}
\end{table*}

Bow shocks are expected around massive runaway stars. Therefore, we used the catalogs of runaway O and Be stars obtained by \cite{MCC2023}, which were identified using \textit{Gaia}~DR3 astrometric data. In this work, the authors cross-matched the GOSC \citep{GOSC} and the BeSS \citep{BeSS} catalogs with \textit{Gaia}~DR3 data. After applying several quality cuts, they produced the GOSC-\textit{Gaia}~DR3 catalog with 417 O-type stars, and the BeSS-\textit{Gaia}~DR3 catalog with 1335 Be-type stars. In the GOSC-\textit{Gaia}~DR3 catalog, they found 106 O runaway stars, with 42 new identifications as runaways. In the BeSS-\textit{Gaia}~DR3 catalog they found 69 Be runaway stars, with 47 new identifications as runaways. The runaway percentages found with respect to the catalogs are 25.4\% and 5.2\% for the O and the Be runaway stars, respectively. The runaway stars were identified through an $E$ value (see Eq.~3 in \citealt{MCC2023}), which indicates the normalized significance of the runaway nature. Therefore, higher $E$ values denote more confident runaway star identifications. The first ten rows of data corresponding to the runaways with the higher $E$ values, of the GOSC and BeSS-\textit{Gaia}~DR3 runaway catalogs, respectively, are shown in Tables~\ref{Tab:GOSC_Runaways}~and~\ref{Tab:BeSS_Runaways}. These tables include the runaway id from the original work, the star names, the \textit{Gaia}~DR3 ids, coordinates (J2000), distances, corrected proper motions (see Sect.~\ref{Sec:Methodology}), $G$ magnitudes, spectral types, the two-dimensional peculiar velocities computed in \cite{MCC2023}, $E$ values, the runaway references (in case the star had already been reported as runaway), and a last column indicating the reference in the cases that they had already been reported to contain bow shocks. As for the runaway id, GR (GOSC Runaways) refers to runaways identified from the GOSC catalog, while BR (BeSS Runaways) to those from the BeSS catalog.

\subsection{Infrared data: WISE}

The Wide-field Infrared Survey Explorer \citep[WISE,][]{WISE2010} was a mission funded by NASA that mapped the whole sky at mid-IR wavelengths from December 2009 to February 2011. It had a much better sensitivity than that of earlier IR survey missions. The different survey bands and resolutions are shown in Table~\ref{Tab:WISE_Survey}.

Bow shocks are related to the warm dust emission, and this emission is mostly expected in the W4 band. Therefore, we used the W4 band to look for bow shock or bubble-like structures around the runaway stars. WISE images are available through the WISE Image Service\footnote{\href{https://irsa.ipac.caltech.edu/applications/wise/}{irsa.ipac.caltech.edu/applications/wise/}} of the NASA/IPAC Infrared Science Archive.

\renewcommand{\arraystretch}{1.1}
\begin{table}
\centering
\caption{Main properties of the WISE survey used in this work.}
\label{Tab:WISE_Survey}
\begin{tabular}{l@{~~~}c@{~~~}c@{~~~}c@{~~~}c@{~~~}}
\hline\hline \vspace{-3mm}\\
Survey & Band & Wavelength   &  Resolution  & Coverage   \\
       &      & ($\mu$m)     &  (\arcsec)   & \\
\hline\vspace{-3mm}\\
WISE   & W1   &  3.4         & 6.1          &  All sky   \\
       & W2   &  4.6         & 6.4          &  All sky   \\
       & W3   &  12          & 6.5          &  All sky   \\
       & W4   &  22          & 12           &  All sky   \\
\hline
\end{tabular}
\end{table}


\section{Methodology} \label{Sec:Methodology}

\subsection{Bow shock searches and identification}

We performed and cross-checked a visual inspection of the W4 images for all the 106 and 69 runaways in Tables~\ref{Tab:GOSC_Runaways}~and~\ref{Tab:BeSS_Runaways}, respectively. We used the astronomical data visualization tool SAOImageDS9 \citep{ds92003} to process the WISE images. We used WISE images of size $20\arcmin\times20\arcmin$ to make sure that we did not lose part of the possible bow shock structures. We looked for bow shock or bubble-like structures around the position of the runaway stars. We also found some features that we would classify not as bow shocks or as bubble-like, but as miscellaneous structures near the runaway stars. These cases will be treated separately. The criterion defined to classify a bow shock or a bubble candidate around a runaway star was based on two aspects:

\begin{enumerate}
    \item The presence of a bow shock or a bubble-like structure around the runaway star.
    \item The arc-shaped structure, or (part of) the rim of the bubble, must be in the direction of the runaway star's corrected proper motion.
\end{enumerate}

To verify the second point, we used the \textit{Gaia}~DR3 proper motions corrected for the ISM motion caused by Galactic rotation. For this, we followed the prescription given in \cite{Comeron2007} (originally from \citealt{SchefflerElsasser1987}), in which it is assumed that the ISM at any given position moves in a circular orbit around the Galactic center (GC) following a rotation curve. This prescription uses the Oort constants (see \citealt{Oort1927,Ogrodnikoff1932}), which can be computed for any given Galactic rotation curve. As was done in \cite{MCC2023}, we used the Galactic rotation curve provided by the A5 fit of \cite{Reid2019} and fit the original data to compute the rotation curve slope. The values we used for the velocity of the Sun, its distance to the GC, the circular rotation velocity at the position of the Sun, and its slope (for distances to the GC larger than 5~kpc) are: $U_{\sun}=10.6 \pm 1.2$~km~s$^{-1}$, $V_{\sun}= 10.7 \pm 6.0$~km~s$^{-1}$, $W_{\sun}=7.6 \pm 0.7$~km~s$^{-1}$, $R_\sun = 8.15 \pm 0.15$~kpc, $\Theta_\sun = 236 \pm 7$~km~s$^{-1}$, and $d\Theta/dR = -1.4 \pm 0.3$~km~s$^{-1}$~kpc$^{-1}$. These are the same values used in all computations presented in \cite{MCC2023}, though some of them are incorrectly quoted in the original manuscript. We could then compute the Oort constants needed to compute the ISM proper motion, and found them to be $A=15.7 \pm 0.5$~km~s$^{-1}$ and $B = - 13.8 \pm 0.5$~km~s$^{-1}$. We finally subtracted the local ISM proper motions to the star's proper motions, and found the proper motions of the star with respect to its local ISM (i.e., the corrected proper motions). The corrected proper motions associated with the runaway stars are listed in Tables~\ref{Tab:GOSC_Runaways}~and~\ref{Tab:BeSS_Runaways}. We also computed the propagation of uncertainties for these variables. From the corrected proper motions and their uncertainties, we computed for each star the position angle ($PA$) from north to east of the corrected proper motions and its uncertainty, $\Delta{PA}$. For the O-runaway stars, $\Delta{PA}$ ranges from 0.7 to 15.5\degr, with a median of 3.9\degr. For the Be-runaway stars, $\Delta{PA}$ ranges from 1.1 to 12.4\degr, with a median of 5.1\degr. Thus, $\Delta{PA}$ is usually only a few degrees; it has a tendency to increase as the peculiar velocity of the star decreases, since the uncertainties of the corrected proper motions are larger compared to the modulus of the corrected proper motion itself. We included the corrected proper motions in $\alpha$, in $\delta$, and its modulus and $PA$, together with their corresponding uncertainties, in the electronic versions of Tables~\ref{Tab:GOSC_Runaways}~and~\ref{Tab:BeSS_Runaways}. Finally, to check for the second item above, we overlapped in the WISE images arrows with 2\arcmin\ length in the direction of the corrected proper motion.

\subsection{Bow shock characterization}

\begin{figure}[hbt!]
\centerline{\includegraphics[width=1\hsize]{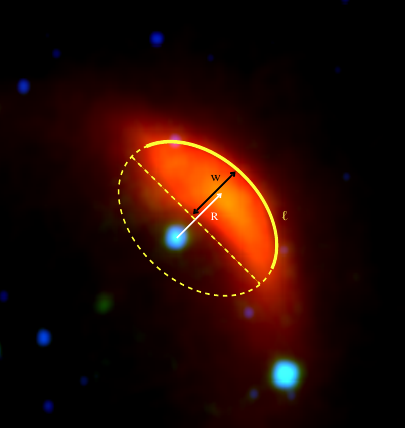}}
\caption{Image of a known bow shock illustrating the geometric characterization used in this work. $R$ is the projected standoff distance, $l$ is the projected length of the bow shock, and $w$ is its projected width. The arc of the ellipse shown with a continuous yellow line was used to derive $l$ (see text for details).}
\label{Fig:Measurements}
\end{figure}

Once the bow shock and bubble-like candidates were identified, we proceeded to their geometrical characterization. We measured the sizes of the IR geometrical structures in the W4 band of the bow shock and bubble candidates projected on the plane of the sky. For this purpose, we used the region tools provided by SAOImageDS9. For the bubble candidates, only the projected width, $w$ -- the diameter of the bubble structure -- was measured. For the bow shock structures, we also measured the projected standoff distance, $R$, which is the distance between the bow shock apex and the star, and the projected length, $l$, of the bow shock. The shape of a bow shock structure can be approximated by a partial ellipse (see Fig.~\ref{Fig:Measurements}). Therefore, we used an ellipse shape to parameterize the bow shock in IR and measure its arc length. For each bow shock, we superimposed an ellipse to the W4 image and obtained its semimajor and semiminor axes. We determined the initial and final angles that define the arc length of the IR emitting region, measured from the center of the ellipse and defined counterclockwise from the semimajor axis. With all these parameters, we can compute the partial perimeter of the ellipse -- that is, the length, $l$ (continuous yellow arc in Fig.~\ref{Fig:Measurements}) -- using the differential arc length equation for a curve defined by an ellipse. We note that these measurements were computed more accurately than the ones presented in \cite{Peri2012} and \cite{Peri2015}. The results of the geometrical measures are presented in Sect.~\ref{Sec:Results}. The values given in linear units (parsec) were computed using the distances provided in Tables~\ref{Tab:GOSC_Runaways}~and~\ref{Tab:BeSS_Runaways}.

\subsection{Interstellar medium ambient density}

We estimated the ISM ambient density, $n_{\text{ISM}}$, at the bow shock position using Eq.~(1) of \cite{Wilkin1996}:
\begin{equation} \label{Eq:nISM}
    R_0 = \sqrt{\frac{\dot{M}\varv_\infty}{4\pi\rho_a {V_\text{PEC}^2}}}~~,
\end{equation}
where $R_0$ is the standoff distance (not the projected one that we measure, $R$), $\dot{M}$ is the stellar mass-loss rate, $\varv_\infty$ is the wind terminal velocity, $V_\text{PEC}$ is the peculiar velocity of the star, and $\rho_a$ is the ambient medium density. From Eq.~\ref{Eq:nISM}, we estimated $\rho_a$, and from $\rho_a = \mu n_{\text{ISM}}$, we finally derived $n_{\text{ISM}}$. In this equation, we assumed that $R_0\sim R$, which will be the case if the motion occurs in the plane of the sky. We note that $n_{\text{ISM}}$ is the volume density in H atoms, and $\mu$ is the mass of the ISM per H atom. We assumed that the fractional abundances for the hydrogen and the helium are $X = 0.9$, and $Y = 0.1$, respectively. Therefore, the mass of the ISM per H atom is then $\mu = \mu_{\text{H}} + 0.1\mu_{\text{He}} = \left[0.9\cdot1.67+0.1\cdot6.6\right]\times 10^{-24}\ \mathrm{g} = 2.16\times 10^{-24}\ \mathrm{g}$.

The stellar parameters, $\dot{M}$ and $\varv_\infty$, for each of our stars were derived from the tables in \cite{Vink2021}\footnote{We also used $\varv_\infty$ from \cite{Prinja1990} as an additional check.}, assuming solar metallicity for our runaway stars, and using the effective temperatures, masses, and luminosities from \cite{Martins2005}. Runaway stars without a luminosity class (GR100 and BR65) were not considered given the large uncertainty in the determination of their temperatures, masses, and luminosities. Given the linear dependence of $\varv_\infty$ on the $T_{\rm eff}$ of the star, we computed a linear regression to the values quoted in \cite{Vink2021}, and used it for our stars. Owing to the variability of the mass-loss rates as a function of $T_{\rm eff}$, we preferred to compute a linear interpolation to assign mass-loss rates to our stars.

For the values of $V_\text{PEC}$, we considered either the 2D or 3D peculiar velocities. The first ones are those from \cite{MCC2023}. To compute the 3D peculiar velocities of the runaway stars, we searched for their heliocentric radial velocities in the literature. For stars with measured values, we computed the regional standard of rest (RSR) velocities in the manner described in \cite{MCC2023}, but using real radial velocities instead of the theoretical ones used there. We then computed the 3D peculiar velocities, $V_\text{PEC}^\text{3D}$, as the moduli of the RSR velocities. We highlight here that we used the 2D and 3D peculiar velocities to compute $n_{\mathrm{ISM}}^{\mathrm{2D}}$ and $n_{\mathrm{ISM}}^{\mathrm{3D}}$, respectively. However, since the 3D peculiar velocities might be contaminated by radial velocity variations in the case of binary systems, $n_{\mathrm{ISM}}^{\mathrm{2D}}$ might be seen as more reliable, although contamination from binarity is unlikely to exceed a few tens of kilometers per second in the case of massive stars. We note that the obtained $n_{\mathrm{ISM}}^{\mathrm{2D}}$ values will always be higher than the $n_{\mathrm{ISM}}^{\mathrm{3D}}$ ones, as lower velocities $\left(V_\text{PEC}^\text{2D} \text{vs.} V_\text{PEC}^\text{3D}\right)$ require higher ISM densities to form a bow shock, assuming all other parameters remain the same in Eq.~\ref{Eq:nISM}.

The stellar parameters, velocities, and computed ISM densities are shown in Tables~\ref{Tab:NewBowShocks}~and~\ref{Tab:KnownBowShocks}. We caveat that the values of $n_{\mathrm{ISM}}$ should be taken as estimates, given the large uncertainties in the determinations of the stellar parameters, and the fact that we used the projected distance, $R$, as the $R_0$ value in Eq.~\ref{Eq:nISM}.


\section{Results} \label{Sec:Results}

During the examination of the W4 images, we found a wide variety of IR structures around the runaway stars: bow shock and bubble-like candidates, intermediate structures\footnote{We call “intermediate” structures those that are halfway between bow shocks and bubbles.}, and miscellaneous structures. There were also some bow shocks or bubbles that were already known. Therefore, we grouped the results in three different categories: new, known, and miscellaneous IR structures. We finish this section by summarizing the search for radio emission from the bow shocks and bubbles discussed in this work.

\subsection{The new candidates for stellar bow shocks or bubbles} \label{Sec:BS_BU_new}

We identified nine new bow shock candidates, three bubble-like candidates, and one intermediate structure for the runaway stars in the GOSC- and BeSS-\textit{Gaia}~DR3 runaway catalogs. These 13 sources are presented in Table.~\ref{Tab:NewBowShocks}. Ten of these structures are associated with O-type runaway stars, and only three of them with Be-type runaway stars. Table~\ref{Tab:NewBowShocks} presents their associated runaway star, with stellar parameters and velocities, the type id (bow shock (BS) or bubble (BU)), and the geometrical measurements, $R$, $l$, and $w$, of the IR structures in the W4 images. We also provide the eccentricity, $e$, of the ellipse used to measure the length of the bow shocks. The last two columns present the estimated $n_{\mathrm{ISM}}$ 2D and 3D values for the bow shock candidates for which we have all the necessary data (see Eq.~\ref{Eq:nISM}, where we used the projected value, $R$, as the real value, $R_0$). For the new bow shock candidates, the 2D ISM density, $n_{\mathrm{ISM}}^{\mathrm{2D}}$, ranges from 0.8 to $\sim$220~cm$^{-3}$, with a median of 6.4~cm$^{-3}$. In the 3D case, $n_{\mathrm{ISM}}^{\mathrm{3D}}$ ranges from 0.6 to $\sim$200~cm$^{-3}$, with 3.8~cm$^{-3}$ as the median\footnote{Using $\varv_\infty$ from \cite{Prinja1990}, we obtained $n_{\text{ISM}}$ values a factor of 1.1--1.8 lower, depending on the case. This also applies to the known bow shocks.}. As is mentioned in Sect.~\ref{Sec:Methodology}, $n_{\mathrm{ISM}}^{\mathrm{2D}}$ values are always higher than the $n_{\mathrm{ISM}}^{\mathrm{3D}}$ ones. BS-GR72 has the largest ISM densities probably due to the high mass-loss rate and wind terminal velocity of its runaway star, combined with a small value for the $R$ distance.

\begin{figure*}[hbt!]\includegraphics[width=\hsize]{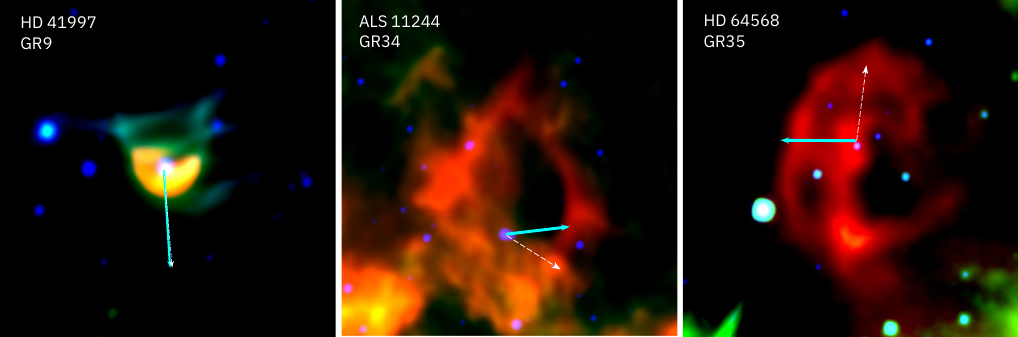}   
\caption{WISE RGB images in W4+W3+W2 for three examples of the new bow shock and bubble candidates presented in Table~\ref{Tab:NewBowShocks} in equatorial coordinates, with north up and east to the left. Dashed white arrows indicate the directions of the original proper motions from \textit{Gaia}~DR3, and solid cyan arrows indicate the directions of the proper motions corrected for the ISM motion caused by Galactic rotation. Each field has a different size, but the arrows are fixed to 2\arcmin. Each panel contains the name of the star provided in the GOSC or the BeSS catalogs, and the runaway star id. The remaining RGB images of the sources from Table~\ref{Tab:NewBowShocks} are presented in Appendix~\ref{Sec:RGBs_New_App}.}
\label{Fig:RGBs_New_MainText}
\end{figure*}

For each candidate, we present the corresponding RGB image in W4+W3+W2 bands in Figs.~\ref{Fig:RGBs_New_MainText}~and~\ref{Fig:RGBs_New_App}. The dashed white arrows indicate the directions of \textit{Gaia}~DR3 proper motions and the solid cyan arrows the directions of the corrected proper motions. We note that in some cases the corrected proper motion direction of the runaway star is not well aligned with the bow shock direction. We tabulate and comment on these differences in Appendix~\ref{Sec:App_DirectionDifferences}. We note that some of these cases could be examples of in situ bow shocks \citep{Povich2008,Kobulnicky2016}, which result from the interaction of outflows from star-forming regions with the stellar wind of massive stars. The orientation of these bow shocks is the result of the flowing ISM and is thus not related to the runaway star motion. Thus, we further inspected the new bow shock fields that showed the largest misalignments, those with angular differences (ADs) greater than $5\sigma$ (see Appendix~\ref{Sec:App_DirectionDifferences}), to determine whether they are in fact in situ bow shocks. For this, we searched for \ion{H}{II} regions or molecular clouds in the Simbad database around the position of these bow shocks. We inspected larger WISE fields centered at the bow shock positions, and finally, we looked for counterparts in \ion{H}{II} region catalogs such as the Green Bank Telescope \ion{H}{II} Region Discovery Survey \citep{Anderson2012} and the WISE Catalog of Galactic \ion{H}{II} Regions \citep{Anderson2014}.

\renewcommand{\arraystretch}{1.1}
\begin{table*}
\centering
\caption{Stellar parameters and velocities of the runaway stars associated with the new bow shock (BS) and bubble (BU) candidates, and geometrical IR measures in the W4 images for the bow shock and bubble candidates.}
\label{Tab:NewBowShocks}
\resizebox{\textwidth}{!}{\begin{tabular}{l@{~}c@{}c@{~~~}c@{~~~}c@{~~~}c@{~~~}c@{~~~}c@{~~~}|@{~}l@{~~~}c@{~~}c@{~~~}c@{~~}c@{~~~}c@{~~}c@{~~~}c@{~}c@{~}c@{~}}
\hline\hline \vspace{-2mm}\\
Runaway Star & ST & $\dot{M}\times10^6$ & $\varv_\infty$ & $\varv_r$ & Ref & $V_\text{PEC}^\text{2D}$ & $V_\text{PEC}^\text{3D}$ & BS/BU  & \multicolumn{2}{c}{$R$} & \multicolumn{2}{c}{$l$~~~~}& \multicolumn{2}{c}{$w$~~~~~~~} & $e$   & $n_{\text{ISM}}^\text{2D}$ & $n_{\text{ISM}}^\text{3D}$
\\  & & (M$_\odot$ yr$^{-1}$)  &(km~s$^{-1}$) &(km~s$^{-1}$) & & (km~s$^{-1}$) & (km~s$^{-1}$) &  &$\left(\arcsec\right)$ & (pc) & $\left(\arcsec\right)$ & (pc) & $\left(\arcsec\right)$   & (pc) & & (cm$^{-3}$) &  (cm$^{-3}$) \\
\hline \vspace{-2mm}\\
  HD 41 997     & O7.5V    & 0.07 & 3349 & $-$18  & 1 & 77.3 & 87.8 & BS-GR9  & 19  & 0.16  & 123 & 1.05        & 38     & 0.32      & 0.51  & 3.6 & 2.8 \\
  ALS 11 244    & O4.5III  & 1.00 & 3465 & $-$39  & 2 & 33.6 & 45.4 & BS-GR34 & 151 & 1.22  & 393 & 3.16         & 56     & 0.45      & 0.37  & 5.1 & 2.8 \\
  HD 64 568     & O3V      & 3.90 & 4175 & 77     & 1 & 40.4 & 42.1 & BS-GR35 & 91  & 1.77  & 309 & 6.02         & 34     & 0.66      & 0.70  & 7.7 & 7.1 \\
  HD 89 137     & ON9.7II  & 0.06 & 3110 & $-$17  & 1 & 39.4 & 45.9 & BS-GR71 & 55  & 0.63  & 171 & 1.96         & 73     & 0.84      & 0.43  & 0.8 & 0.6 \\
  BD $+$50 886  & O4V      & 4.01 & 3984 & $-$36  & 2 & 30.5 & 32.2 & BS-GR72 & 30  & 0.43  & 223 & 3.17         & 30     & 0.43      & 0.18  & 221.2 & 198.6\\
  HDE 338 916   & O7.5V    & 0.07 & 3349 & $-$1.5 & 3 & 21.2 & 23.2 & BS-GR75 & 28  & 0.28  & 310 & 3.11         & 110    & 1.10      & 0.47  & 15.2 & 12.7\\
  HD 192 001    & O9.5IV   & 0.07 & 3157 & $-$27  & 1 & 21.7 & 28.7 & BU-GR78 & --  & --    &  --   &  --    & 125    & 1.03      & --    & --  & -- \\  
  HD 189 957    & O9.7III  & 0.07 & 3116 & 39.9   & 1 & 21.0 & 52.1 & BS-GR80 & 20  & 0.19  & 178 & 1.74         & 90     & 0.88      & 0.29  & 29.9& 4.9  \\
  HD 116 282    & O8III   & 0.05 & 3268 & $-$39  & 3 & 28.0 & 28.6 & BS-GR93 & 32  & 0.60  & 225 & 4.21         & 101    & 1.89      & 0.21  & 1.4 & 1.4  \\
  ALS 17 591    & O5:      &  --  & --   &$-$10.2 & 3 & 16.0 & 49.8 & BU/BS-GR100 & --/13 &--/0.20 & --/261& --/4.13 & 69/74  & 1.09/1.17 & --/0.51 &  -- & -- \\  
\hline \vspace{-3mm}\\
  V372 Sge         & B0.5eIII  & 0.05 & 3044 & $-$18 & 1  & 44.4 & 51.6 & BU-BR21   &  -- & --   &  --  & --   & 105 & 0.99 & --   & --& --\\
  HD 225985        & B1psheV   & 0.06 & 3053 & --    &--  & 32.9 & --   & BU-BR46   & --  & --   &  --  & --   & 78  & 0.67 & --   & --& --\\
  Cl* NGC 663 G 159& Be        & --   & --   & --    &--  & 30.0 & --   & BS-BR65   & 112 & 0.26 & 138 & 0.32 & 36   & 0.08 & 0.71 & --& --\\
\hline
\end{tabular}}
\tablefoot{The left side of the table provides information associated with the runaway stars, while the right side provides information associated with the corresponding bow shock and bubble candidates. Spectral types (STs) were taken from the GOSC \citep{GOSC} and the BeSS \citep{BeSS} catalogs. $\dot{M}$, mass-loss rates, were linearly interpolated from \cite{Vink2021}. Wind terminal velocities were obtained by applying a linear regression to \cite{Vink2021} data. Radial velocity references: (1)~\citet{Bruijne2012}; (2)~\citet{Kobulnicky2022}; (3)~\citet{Kharchenko2007}. $V_\text{PEC}^\text{2D}$ are those from \cite{MCC2023} and 
$V_\text{PEC}^\text{3D}$ are obtained with the same methodology using the quoted $\varv_r$ values (see text for details). The right side of the table provides the classification id of the IR structure around the runaway star: BS for bow shocks; BU for bubbles; and BU/BS and BS/BU for intermediate structures between BS and BU. $R$ is the projected standoff distance from the star to the midpoint of the bow shock, $l$ is the length, and $w$ is the width of the bow shock structure. These distances are projected in the plane of the sky. For bubbles, only the width was measured. Column $e$ provides the eccentricity of the ellipse used to measure the bow shocks. $n_{\text{ISM}}^\text{2D}$ and $n_{\text{ISM}}^\text{3D}$ are the ISM densities of the medium around the bow shock position considering either $V_\text{PEC}^\text{2D}$ or $V_\text{PEC}^\text{3D}$ as $V_\text{PEC}$ in Eq.~\ref{Eq:nISM}.} The horizontal line separates O-type stars (upper part) from Be-type stars (bottom part). This table is available at the CDS.
\end{table*}

We shall comment now on the three cases with AD above $5\sigma$: BS-GR71, BS-GR72, and BS-GR93. For BS-GR71, there were no \ion{H}{II} regions found on the field. Therefore, this cannot be a case of an in situ bow shock and we have kept it as a bow shock candidate associated with the runaway star GR71. In contrast, BS-GR72 is coincident with the evolved \ion{H}{II} region \object{S206} (or \object{NGC~1491}), located at a distance of $2.96^{+0.17}_{-0.15}$~kpc \citep{MendezDelgado2022}. This distance is compatible with the one of the runaway star GR72, $2.94\pm0.3$~kpc \citep{MCC2023}. Therefore, this bow shock could be indeed an in situ bow shock. In addition, we note that this \ion{H}{II} region \object{S206} presents radio emission (see Appendix~\ref{Sec:App_Radio}). BS-GR93 has the molecular cloud \object{[DB2002b] G307.25+2.86} \citep{Hartley1986} at $\sim$10\arcmin, related to the globule \object{BHR 87}, whose distance appears to have different values in the literature: 400~pc \citep{Launhardt2010} and 1~kpc \citep{Bourke2005}. However, none of these distances is compatible with the one of the runaway star GR93, $3.85\pm0.35$~kpc. Therefore, we cannot claim that BS-GR93 is an in situ bow shock candidate.

\begin{figure}[hbt!]
\centerline{\includegraphics[width=0.85\hsize]{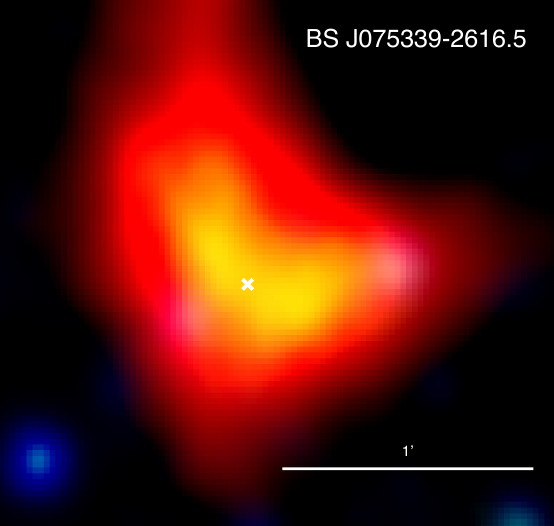}}
\caption{WISE RGB image in W4+W3+W2 for the bow shock candidate \object{BS~J075339$-$2616.5} in equatorial coordinates, with north up and east to the left. The white cross indicates the center of the bow shock emission, at coordinates RA = 07h 53m 39.1s and DEC = $-$26\degr\ 16\arcmin\ 31\arcsec. This bow shock appeared in the WISE field of BS-GR35 (see Fig.~\ref{Fig:RGBs_New_MainText}), and has no associated runaway star.}
\label{Fig:RGB_BSJ075339-2616.5}
\end{figure}

We shall now comment on two particularly interesting cases, BS-GR35, and BU-BR46. While working with the RGB image of BS-GR35, we discovered another bow shock candidate in the field, $\sim$2\arcmin\ south from the runaway GR35 (see Fig.~\ref{Fig:RGBs_New_MainText}). We called it \object{BS~J075339$-$2616.5} because of the coordinates of its center: RA = 07h 53m 39.1s and DEC = $-$26\degr\ 16\arcmin\ 31\arcsec. It has a perfect bow structure in the W4 and W3 images. We show an RGB image centered and zoomed on this source in Fig.~\ref{Fig:RGB_BSJ075339-2616.5}. We note that it cannot be associated with our runaway star GR35 because of its position and orientation with respect to the star and its proper motions. We looked for possible \textit{Gaia} stars nearby \object{BS~J075339$-$2616.5} but did not find a clear candidate. We also looked for possible associated sources in the region, but there is only one IRAS source, \object{IRAS~07515$-$2608}. Notably, we found a radio detection for this new bow shock candidate in RACS-low and RACS-mid images, but it does not appear to be a clear counterpart (see Appendix~\ref{Sec:App_Radiosearch}). 
As for BU-BR46, we originally proposed it as a bow shock candidate. However, after a detailed examination of the IR structure, we classified it as a bubble. We found that the bow shape was actually caused by another star inside the W4 region of emission, to the southwest of BU-BR46 (see Fig.~\ref{Fig:RGBs_New_App}). It is a long-period variable star, \object{2MASS~J19493059+3257068}, that is not physically related to our runaway.

\subsection{Known candidates for stellar bow shocks or bubbles}

There are a total of 17 known bow shocks or bubble-like candidates among the GOSC-\textit{Gaia}~DR3 runaway stars. There is only one known candidate among the BeSS-\textit{Gaia}~DR3 runaways, BS-BR14, which is indeed coincident with BS-GR42 because it is a runaway star common to the two catalogs\footnote{This star is classified as a O9.2 type star in GOSC, while it appears as O9e in BeSS.}. Their bow shock or bubble references are from different catalog searches \citep{Peri2012,Peri2015,Kobulnicky2016,MA2018,Moutzouri2022}, and are included in Tables~\ref{Tab:GOSC_Runaways}~and~\ref{Tab:BeSS_Runaways}. However, some of them have not been geometrically characterized in previous works. In this section, we focus on these cases. We selected those bow shocks and characterized them using the same procedure as for the new bow shocks discovered in this work. Table~\ref{Tab:KnownBowShocks} presents their associated runaway stars, with stellar parameters and velocities, the type id, the geometrical measurements of the IR structures in W4, and the eccentricity, $e$. Again, we estimated the ISM densities for the known bow shock candidates. In this case, $n_{\mathrm{ISM}}^{\mathrm{2D}}$ ranges from 0.4 to $\sim$200~cm$^{-3}$, and $n_{\mathrm{ISM}}^{\mathrm{3D}}$ ranges from 0.3 to $\sim$190~cm$^{-3}$, with BS-GR83 presenting the largest value due to its short $R$ distance. 

\begin{figure*}[h!]
 
\includegraphics[width=\hsize]{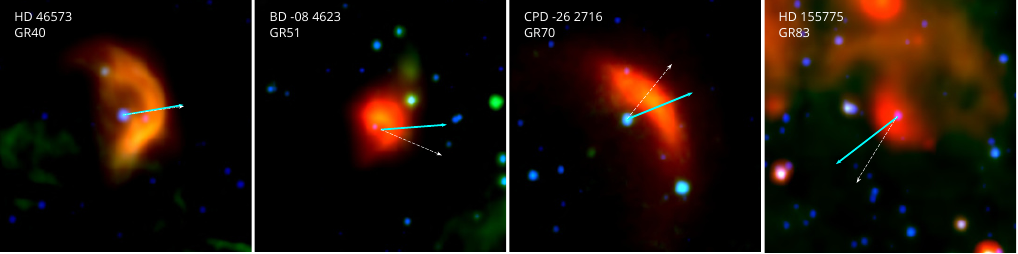}

\caption{WISE RGB images in W4+W3+W2 for the known bow shock and bubble candidates presented in Table~\ref{Tab:KnownBowShocks} in equatorial coordinates, with north up and east to the left. Dashed white arrows indicate the directions of the original proper motions from \textit{Gaia}~DR3, and solid cyan arrows indicate the directions of the proper motions corrected for the ISM motion caused by Galactic rotation. Each field has a different size, but the arrows are fixed to 2\arcmin. Each panel contains the name of the star as provided in the GOSC or the BeSS catalogs, and the runaway star id.}
\label{Fig:RGBs_Known}
\end{figure*}

As for the new candidates, we also present the corresponding WISE RGB images in W4+W3+W2 bands in Fig.~\ref{Fig:RGBs_Known}. We shall comment here on two special cases: BS/BU-GR51 and EB43 of \cite{Peri2015}, coincident with GR20.

\renewcommand{\arraystretch}{1.1}
\begin{table*}
\centering
\caption{Stellar parameters and velocities of the runaway stars associated with the known bow shock (BS) and bubble (BU) candidates that were not characterized by previous works, and geometrical IR measures in the W4 images for the bow shock and bubble candidates.}
\label{Tab:KnownBowShocks}
\resizebox{\textwidth}{!}{\begin{tabular}{l@{~}c@{}c@{~~~}c@{~~~}c@{~~~}c@{~~~}c@{~~~}c@{~~~}|l@{~~~}c@{~~}c@{~~~}c@{~~}c@{~~~}c@{~~}c@{~~~}c@{~}c@{~}c@{~}}
\hline\hline \vspace{-2mm}\\
Runaway Star & ST & $\dot{M}\times10^6$ & $\varv_\infty$ & $\varv_r$ & Ref & $V_\text{PEC}^\text{2D}$ & $V_\text{PEC}^\text{3D}$ & BS/BU  & \multicolumn{2}{c}{$R$} & \multicolumn{2}{c}{$l$~~~~}& \multicolumn{2}{c}{$w$~~~~~~~} & $e$   & $n_{\text{ISM}}^\text{2D}$  & $n_{\text{ISM}}^\text{3D}$
\\  & & (M$_\odot$ yr$^{-1}$)  &(km~s$^{-1}$) &(km~s$^{-1}$) & &  
(km~s$^{-1}$) & (km~s$^{-1}$) & &$\left(\arcsec\right)$ & (pc) & $\left(\arcsec\right)$ & (pc) & $\left(\arcsec\right)$   & (pc) & & (cm$^{-3}$) & (cm$^{-3}$)\\
\hline \vspace{-2mm}\\

HD 46 573  & O7V & 0.11 & 2992  & ~29.3 & 1 & 40.4  & 40.4  & BS-GR40  & 54    & 0.35  & 369    & 2.40    & 88      & 0.57      & 0.54  & 3.8 & ~~3.8 \\

BD $-$08 4623 &  B0.5:Ia: & 0.50 & 2293 & -- & -- & 28.8   & -- & BS/BU-GR51\tablefootmark{*}  & 21/-- & 0.22/-- & 200/-- & 2.05/-- & 74/74   & 0.76/0.76 & 0.46/-- & 71.5 & --/--  \\

CPD $-$26 2716 & O6.5Iab & 0.60 & 2909  & ~57.8 & 2  & 73.7 & 77.7  & BS-GR70  & 53 & 1.45 & 292 & 7.97 & 54          & 1.47 & 0.51  & 0.4 & ~~~0.3 \\
 
HD 155 775  & O9.7III & 0.07 & 3116  & $-$9.0 & 1 & 23.8 & 24.4  & BS-GR83  & 15 & 0.07 & 118 & 0.54 & 56   & 0.26 & 0.54 & 202.5 & 192.6\\

\hline
\end{tabular}}
\tablefoot{The left side of the table provides information associated with the runaway stars, while the right side provides information associated with the corresponding bow shock and bubble candidates. Spectral types (STs) were taken from the GOSC \citep{GOSC} and the BeSS \citep{BeSS} catalogs. $\dot{M}$, mass-loss rates, were linearly interpolated from \cite{Vink2021}. Wind terminal velocities were obtained by applying a linear regression to \cite{Vink2021} data. Radial velocity references: (1)~\citet{Bruijne2012}; (2)~\citet{Williams2011}. $V_\text{PEC}^\text{2D}$ are those from \cite{MCC2023} and 
$V_\text{PEC}^\text{3D}$ were obtained with the same methodology, using the quoted $\varv_r$ values (see text for details). 
The right side of the table provides the classification id of the IR structure around the runaway star: BS for bow shocks; BU for bubbles; and BU/BS and BS/BU for intermediate structures between BS and BU. $R$ is the projected standoff distance from the star to the midpoint of the bow shock, $l$ is the length, and $w$ is the width of the bow shock structure. These distances are projected in the plane of the sky. For bubbles, only the width was measured. Column $e$ provides the eccentricity of the ellipse used to measure the bow shocks. $n_{\text{ISM}}^\text{2D}$ and $n_{\text{ISM}}^\text{3D}$ are the ISM densities of the medium around the bow shock position considering either $V_\text{PEC}^\text{2D}$ or $V_\text{PEC}^\text{3D}$ as $V_\text{PEC}$ in Eq.~\ref{Eq:nISM}.
This table is available at the CDS.\\
\tablefoottext{*}{Previous works classified it as bubble; we classified it as an intermediate structure.}
}
\end{table*}

BS/BU-GR51 is a known IR structure identified in \cite{MA2018}. In that work, they introduce it as an asymmetric region of hot dust around the runaway star rather than a bow shock. The likely existence of an IR bubble around GR51 was already identified in \cite{Simpson2012}. However, given the asymmetric morphology, we have classified this candidate as an intermediate structure, since a bubble should be more spherical. The RGB image in W4+W3+W2 can be seen in Fig.~\ref{Fig:RGBs_Known}.

We also comment on the particular case of EB43, a bow shock candidate of \cite{Peri2015}. The position of this bow shock is coincident to the runaway GR20. However, the strange shape of EB43, cut in half just in the direction in which the bow shock is expected, does not meet the criterion presented in Sect.~\ref{Sec:Methodology} and has made us discard it as a bow shock candidate. We note that it was already proposed as a doubtful candidate on the basis of morphology in \cite{Peri2015}. Therefore, although the number of known bow shocks within our sample was initially 17, after discarding this one we considered only 16 cases.

\subsection{Miscellaneous infrared structures} \label{Sec:Miscellaneous}

In this section, we present different types of IR structures found in the W4 images around the runaway stars. We cannot classify these various structures as bow shock or bubble candidates with certainty, but they present a particular morphology that makes them interesting in their own right. Therefore, we decided to group them in a category we called miscellaneous IR structures, and dedicate this section to them. All the sources belonging to this category are presented in Table~\ref{Tab:Miscellaneous}, along with a morphological classification. In what follows, we comment on a particular subgroup that we call mini-bubbles and discuss the possible origin of the W4 emission.

When examining the W4 images, we found some runaway stars that exhibit relatively compact emission: 35 and 22 sources among the GOSC- and the BeSS-\textit{Gaia}~DR3 runaways, respectively. We note that all of them look point-like in the other WISE bands. We applied Guassian fits to the W4 images at the position of these sources. We found that most of them are compatible with being point-like. However, we found two particular Be-runaway stars, BR10 and BR15, that have excesses spanning around two times the WISE PSF in W4 (which is 12\arcsec). We then classified the objects in this subgroup as mini-bubbles, since they are not as extended and big as typical bubbles (the minimum size among our bubbles is 69\arcsec), but they are not compatible with point-like source emission\footnote{We inspected the availability of 24~$\mu$m Spitzer images for the point-like sources and mini-bubbles listed in Table~\ref{Tab:Miscellaneous}. We only found data for GR84, which is point-like as in the W4 WISE image.}.

We looked for explanations for the W4 excesses found in point-like sources and mini-bubbles in the literature, but did not find a clear pattern. In the case of Be-type stars, this IR emission could be related to the presence of a circumstellar disk \citep{Rivinius2013}. We found a runaway star in the literature, \object{V452 Sct}, with an IR excess in W4 that is visually consistent with what we have called mini-bubbles. Its RGB image is presented in Fig.~6 of \cite{MA2018}. In this work, they present the IR emission as an excess in W4 consistent with it being a point source centered on the star, but not a bow shock. They also found that this excess was studied in \cite{Miroshnichenko2000}, which proposed that it was likely caused by an LBV-type outburst in the past. We applied a Guassian fit to the W4 excess, and found that the size was larger than twice the WISE PSF in W4. Therefore, we also consider this runaway star to be a mini-bubble. 

We also inspected the point-like sources and mini-bubbles with the VizieR Photometry viewer tool, which allows for easy visualization of photometry-enabled catalogs in VizieR. We used this tool to search for components in addition to the expected blackbody stellar component. For the case of V452~Sct, the mini-bubble of \cite{MA2018}, two components can be clearly distinguished. However, we only found another clear bump in BR29, and some hints in GR41, BR10, BR15, BR28, and BR34. Therefore, it is difficult to associate the mini-bubble nature with any additional component in their SEDs. We present in Appendix~\ref{Sec:App_Miscellaneous} the RGB images for the seven miscellaneous sources from Table~\ref{Tab:Miscellaneous} that are not point-like.

\renewcommand{\arraystretch}{1.1}
\begin{table*}
\centering
\caption{Miscellaneous IR structures and their W4 morphology.}
\label{Tab:Miscellaneous}
\resizebox{\textwidth}{!}{\begin{tabular}{l@{~~~}c@{~~~}|l@{~~~}c@{~~~}|l@{~~~}c@{~~~}|l@{~~~}c@{~~~}|l@{~~~}c@{~~~}|l@{~~~}c@{~~~}}
\hline\hline \vspace{-2mm}\\
Run. id & W4 Morph. & Run. id & W4 Morph. & Run. id & W4 Morph. & Run. id & W4 Morph. & Run. id & W4 Morph. & Run. id & W4 Morph.\\
\hline \vspace{-2mm}\\
  GR4     & $\bullet$  & GR6    & $\bullet$    & GR10   & $\bullet$  &
  GR11    & $\bullet$  & GR19   & $\bullet$    & GR21   & $\bullet$ \\  
  GR23    & $\bullet$  & GR30   & $\bullet$    & GR32   & $\bullet$ &
  GR33    & $\bullet$  & GR36   & $\bullet$    & GR37   & $\bullet$ \\
  GR38    & $\supset$  & GR41   & $\bullet$    & GR43   & $\bullet$ &
  GR46    & $\bullet$  & GR47   & $\supset$    & GR48   & $\bullet$ \\
  GR49    & $\bullet$  & GR53   & $\bullet$    & GR55   & $\bullet$ &
  GR57    & $\bullet$  & GR58   & $\bullet$    & GR60   & $\bullet$ \\
  GR63    & $\bullet$  & GR64   & $\bullet$    & GR66   & $\bullet$ &
  GR68    & $\bigcirc$ & GR69   & $\bullet$    & GR74   & $\bullet$ \\
  GR79    & $\bullet$  & GR81   & $\bigcirc$   & GR84   & $\bullet$ &
  GR85    & $\bullet$  & GR86   & $\bullet$    & GR87   & $\bullet$ \\
  GR94    & $\bullet$  & GR96   & $\bullet$    & GR98   & $\bullet$ &
  GR101   & $\supset$  & & & &\\ 
\hline
  BR1     & $\bullet$  & BR2    & $\bullet$    & BR3    & $\bullet$ &
  BR4     & $\bullet$  & BR5    & $\bullet$    & BR7    & $\bullet$ \\
  BR8     & $\bullet$  & BR9    & $\bullet$    & BR10   & $\circ$   &
  BR11    & $\bullet$  & BR15   & $\circ$      & BR16   & $\bullet$ \\
  BR20    & $\bullet$  & BR27   & $\bullet$    & BR28   & $\bullet$ &
  BR29    & $\bullet$  & BR30   & $\bullet$    & BR32   & $\bullet$ \\
  BR34    & $\bullet$  & BR37   & $\bullet$    & BR45   & $\bullet$ &
  BR58    & $\bullet$  &        &              &        & \\
\hline
\end{tabular}}
\tablefoot{Morphology observed in W4 image: $\supset$ curved structure; $\bigcirc$ kind of bubble; $\bullet$ point-like source; $\circ$ mini-bubble.}
\end{table*}

\subsection{Search for radio counterparts} \label{Sec:Radio}

We conducted a search for radio emission among the newly detected IR structures. Details of this search can be found in Appendix~\ref{Sec:App_Radio}, with data presented in Appendix~\ref{Sec:App_Radiodata} and the obtained results shown in Appendix~\ref{Sec:App_Radiosearch}. We did not find point-like radio emission coincident with the positions of the runaway stars. We did not find any clear radio counterparts of the IR sources either, but we found radio emission partially overlapping them in six particular cases that are discussed in Appendix~\ref{Sec:App_Radiosearch}: BS-GR9, BS-GR72, BS-GR75, \object{BS~J075339$-$2616.5}, BS/BU-GR51, and BS-GR83.

We also investigated a thermal or nonthermal origin of the possible radio emission for the bow shock candidates, based on their IR characterization and the RACS radio data used in this work (see Appendix~\ref{Sec:App_Emission_mechanisms}). We conclude that the sources BS-GR75 and BS-GR83 should be detectable in radio for standard parameters. Future, more sensitive radio observations could reveal radio emission in these sources, which are more likely to show nonthermal than thermal radio emission (see details in Appendix~\ref{Sec:App_Emission_mechanisms}).

\section{Discussion} \label{Sec:Discussion}

In this section, we compare our results with those of previous works in terms of bow shock searches, the geometrical characterization of the IR structures, and the estimated ISM densities. We found nine new stellar bow shock candidates, three new bubble candidates, and one new intermediate structure candidate within the catalogs of O and Be-runaway stars compiled in \cite{MCC2023}. Among the catalogs, there were also 17 known bow shock candidates (see Bow. Ref. column in Tables~\ref{Tab:GOSC_Runaways}~and~\ref{Tab:BeSS_Runaways}). The mean $V_\text{PEC}^\text{2D}$ are 34~km~s$^{-1}$, and 42~km~s$^{-1}$, for the runaways associated with the new and the known IR structures, respectively. The mean $E$ values are 1.53 and 1.84 for the runaway stars associated with the new and the known IR structures, respectively. The runaways associated with the known bow shocks have higher velocities and $E$ values because they were the most obvious and easiest to find. In \cite{MCC2023}, we discover new runaways that are closer to the runaway threshold ($E\sim1$, not as evident as the already known ones), and that thus have slower velocities. More than half of our new discoveries in the IR are associated with the new runaway discoveries in \cite{MCC2023}, and they are closest to the runaway threshold. Therefore, more detailed searches for runaway stars, such as the one presented in \cite{MCC2023}, working with space velocities and the accurate \textit{Gaia}~DR3 data, open the door to identifying both new runaway stars and bow shock structures around them. 

Among the 17 known bow shocks found, we discarded the one associated with GR20 (EB43 in \citealt{Peri2015}), and reclassified the bubble of GR51 as an intermediate structure. In this work, we thus confirm 16 known candidates as bow shocks. The known bow shocks were identified for the first time in the following works: \cite{Peri2012}, \cite{Peri2015}, \cite{Kobulnicky2016}, \cite{MA2018}, and \cite{Moutzouri2022}.

We note that among the 106 O-type runaway stars we have found 9 new bow shocks plus 16 known ones, representing 23.6\%. This is very similar to the percentage one can obtain from \cite{Peri2015} considering the sample of O-type runaways of \cite{Tetzlaff2011}: 11 bow shocks among 54 runaways representing 20.4\%. In contrast, for Be-type runaway stars we obtain 2 bow shocks among the 69 runaway stars in our sample, representing $\sim$3\%. By comparison, among the 12 Be runaway stars from \cite{Tetzlaff2011}, there were no bow shocks identified by \cite{Peri2015}.

From the references listed in the previous paragraph, only \cite{Peri2015} and \cite{Moutzouri2022} provide a geometrical characterization of the bow shocks in W4-band WISE images. In this work, we also include the geometrical measurements for both the bow shock and the bubble candidates. We also characterize those known bow shocks of our runaway catalogs that are not geometrically characterized in \cite{Kobulnicky2016}, \cite{MA2018}, \cite{Bodensteiner2018}, and \cite{Jayasinghe2019}. The order of magnitude for $R$ and $w$ is less than 1~pc, while for $l$ it is a few parsecs. We conclude that both new and known bow shocks have similar and compatible $R$, $l$, and $w$ to those presented in Peri et al. However, in this work we have improved the procedure of measurement, using the SAOImageDS9 software to measure $l$ through ellipse shapes. We have also provided the eccentricity of the ellipses used to measure the bow shocks. We highlight the importance of the geometrical characterization of these objects, as it can be used to put constraints on the underlying radio emission mechanism of these sources \citep{VandenEijnden2022b}.

The surrounding ISM densities, $n_{\mathrm{ISM}}^{\mathrm{2D}}$ and $n_{\mathrm{ISM}}^{\mathrm{3D}}$, were estimated for 2D and 3D peculiar velocities, respectively, using Eq.~\ref{Eq:nISM}, and are presented in Tables~\ref{Tab:NewBowShocks}~and~\ref{Tab:KnownBowShocks}. The medians obtained for $n_{\mathrm{ISM}}^{\mathrm{2D}}$ and $n_{\mathrm{ISM}}^{\mathrm{3D}}$ are 6.4 and 3.8~cm$^{-3}$, respectively. We note that these values were computed with our reduced sample of 12 bow shocks, for which we could estimate ISM densities. In \cite{Peri2012} and \cite{Peri2015}, they computed $n_{\text{ISM}}$ in 3D for 28 and 31 sources, respectively. The median $n_{\text{ISM}}$ for all their 59 sources is $\sim$2~cm$^{-3}$. Despite our reduced sample, our 3D median and theirs are quite similar. Their sample is much larger, so their median should be more reliable than ours. However, we remark that our geometrical measures are more accurate than theirs (see Sect.~\ref{Sec:Methodology}). The typical densities of the cold neutral ISM have a wide dispersion, in a range from 1--100~cm$^{-3}$ (see \citealt{Kulkarni1988} for a review and the introduction of the recent paper \citealt{PatraRoy2024}). Our values fit well within this range, with two particular exceptions with densities of $\simeq 200$~cm$^{-3}$, BS-GR72 and BS-GR83. We note that small $R$ measures result in large $n_{\text{ISM}}$ estimates. Therefore, projection effects may overestimate ISM densities. This is probably the case for BS-GR83. However, for BS-GR72 it seems to be a combination between its small $R$, high mass-loss rate, and wind terminal velocity. However, if BS-GR72 were an in situ bow shock, the $n_{\text{ISM}}$ estimate would not be so relevant in this particular case. We emphasize again that the individual results in ISM density estimates should be used with caution because they are derived from a combination of $\dot{M}$, $\varv_\infty$, $V_\text{PEC}^\text{2D}$ or $V_\text{PEC}^\text{3D}$, and $R$ values, for which: i) $R$ is used instead of $R_0$ in Eq.~\ref{Eq:nISM}, though in particular cases they might differ significantly for very small values of $R$, and ii) it is difficult to obtain accurate estimates of $\dot{M}$ and $\varv_\infty$ values.

\section{Summary and future work} \label{Sec:Summary}

In this work, we have searched for stellar bow shocks around the new sample of O and Be runaway stars provided in \cite{MCC2023}, using WISE IR images, \textit{Gaia}~DR3 astrometric data, and radio data. The main conclusions of these work are summarized below.

While searching for stellar bow shocks around runaway stars, we found nine new stellar bow shock candidates, three new bubble candidates, and one new intermediate structure. Of these, ten are associated with O-type stars, and only three with Be-type stars. We also visually found a bow shock candidate in the BS-GR35 field, with no associated runaway star, which we called \object{BS~J075339$-$2616.5}. We investigated and quantified the possible misalignment between the runaway star corrected proper motions and bow shock directions, and we found BS-GR72 as an in situ bow shock candidate related the \ion{H}{II} region \object{S206}. We identified 17 known bow shocks, all of which are associated with O-type stars, but discarded the one associated with GR20 \citep[EB43 in][]{Peri2015}. We identified 62 miscellaneous IR structures, 40 and 22 of which are associated with O- and Be-type stars, respectively. Among them, we introduced a new designation for bubble-like sources with around twice the W4 PSF: mini-bubble sources. We found two, BR10 and BR15, which are Be-type stars. We include under this new designation sources such as V452 Sct, of \cite{MA2018}. More detailed studies of these sources could reveal the nature of the IR emission around these runaways.

On the other hand, we conducted a geometrical characterization of the sources through IR measures in W4. For bow shocks, we measured the projected standoff distance, $R$, length, $l$, and width, $w$, while for bubbles we only provided the width. The values of $R$ and $w$ are always similar, and their order of magnitude is less than one parsec. The values of $l$ are typically a few parsecs. The geometrical IR measures, together with mass-loss rates and stellar wind velocities of the runaway stars, allowed us to estimate the ISM density of the medium in a 2D- and 3D-peculiar velocity case. We found $n_{\mathrm{ISM}}^{\mathrm{2D}}$ in the range from 0.8 to $\sim$220~cm$^{-3}$, with a median of 6.4~cm$^{-3}$, and $n_{\mathrm{ISM}}^{\mathrm{3D}}$ in the range from 0.3 to $\sim$200~cm$^{-3}$, with a median of 3.8~cm$^{-3}$. Particularly high values could be overestimated due to projection effects in the $R$ measurement. Both the geometrical IR measures and the $n_{\text{ISM}}$ estimates obtained are in agreement with previous works (\citealt{Peri2012, Peri2015}; \citealt{PatraRoy2024} and references therein). We note that the obtained values of $n_{\text{ISM}}$, even considering the limitations outlined in Sect.~\ref{Sec:Discussion}, are estimated at the position of our sources. Therefore, they could be useful for other studies at those positions, such as stellar formation studies or radiation processes in these regions.

The extensive bow shock and bubble searches performed in works such as \cite{Peri2012}, \cite{Peri2015}, \cite{Kobulnicky2016}, \cite{Bodensteiner2018}, and \cite{Jayasinghe2019} leave few options for finding new ones from such sources in the sky. However, deeper searches for runaway stars using accurate \textit{Gaia}~DR3 data \citep{MCC2023} allow new runaway stars to be found closer to runaway thresholds, and then a still-hidden population of bow shocks. Most of the new bow shock candidates identified correspond to new runaway discoveries in \cite{MCC2023}. They also have smaller $V_\text{PEC}^\text{2D}$ velocities than known bow shocks. The differences are, on average, about $\sim$~7~km~s$^{-1}$. As for the bow shock percentage in runaways by spectral type, we obtained 23.6\% and $\sim$3\% for O and Be runaway stars, respectively. On the other hand, the main difference with previous works of general bow shock searches is that here we provide the geometrical IR characterization of the sources identified. These measurements are very useful to assess the underlying radio emission mechanism in these sources (see \citealt{VandenEijnden2022b}).

In this context, we searched for radio emission from these bow shocks in archival radio data and we evaluated the possible thermal or nonthermal origin of the radio emission mechanisms for our sources. More sensitive observations are needed to detect bow shocks at radio wavelengths, uncover a previously undetected population of radio-emitting bow shocks, and shed light on their radio and HE emission connection.

\begin{acknowledgements}

       We thank the anonymous referee for useful suggestions and comments that helped to improve the content of the manuscript.
       This work made used of the public JUPYTER notebook of \cite{VandenEijnden2022b} to reproduce Fig.~\ref{Fig:PredictedRadioFlux}. We thank the authors of that work for the data availability. We thank T. Antoja for useful discussions. We acknowledge financial support from the State Agency for Research of the Spanish Ministry of Science and Innovation under grants PID2019-105510GB-C31/AEI/10.13039/501100011033, PID2019-104114RB-C33/AEI/10.13039/501100011033, PID2022-136828NB-C41/AEI/10.13039/501100011033/ERDF/EU, and PID2022-138172NB-C43/AEI/10.13039/501100011033/ERDF/EU, and through the Unit of Excellence María de Maeztu 2020-2023 award to the Institute of Cosmos Sciences (CEX2019-000918-M). We acknowledge financial support from Departament de Recerca i Universitats of Generalitat de Catalunya through grant 2021SGR00679. MC-C acknowledges the grant PRE2020-094140 funded by MCIN/AEI/10.13039/501100011033 and FSE/ESF funds.
      This publication makes use of data products from the Wide-field Infrared Survey Explorer, which is a joint project of the University of California, Los Angeles, and the Jet Propulsion Laboratory/California Institute of Technology, funded by the National Aeronautics and Space Administration. This work has made use of data from the European Space Agency (ESA) mission {\it Gaia} (\url{https://www.cosmos.esa.int/gaia}), processed by the {\it Gaia} Data Processing and Analysis Consortium (DPAC, \url{https://www.cosmos.esa.int/web/gaia/dpac/consortium}). Funding for the DPAC has been provided by national institutions, in particular the institutions participating in the {\it Gaia} Multilateral Agreement. This research has made use of NASA’s Astrophysics Data System.
      This research has made use of the SIMBAD database, operated at CDS, Strasbourg, France.
      
\end{acknowledgements}

\bibliographystyle{aa} 
\bibliography{mybibliography} 

\begin{thebibliography}{79}
\expandafter\ifx\csname natexlab\endcsname\relax\def\natexlab#1{#1}\fi

\bibitem[{{Abdollahi} {et~al.}(2022){Abdollahi}, {Acero}, {Baldini}, {Ballet}, {Bastieri}, {Bellazzini}, {Berenji}, {Berretta}, {Bissaldi}, {Blandford}, {Bloom}, {Bonino}, {Brill}, {Britto}, {Bruel}, {Burnett}, {Buson}, {Cameron}, {Caputo}, {Caraveo}, {Castro}, {Chaty}, {Cheung}, {Chiaro}, {Cibrario}, {Ciprini}, {Coronado-Bl{\'a}zquez}, {Crnogorcevic}, {Cutini}, {D'Ammando}, {De Gaetano}, {Digel}, {Di Lalla}, {Dirirsa}, {Di Venere}, {Dom{\'\i}nguez}, {Fallah Ramazani}, {Fegan}, {Ferrara}, {Fiori}, {Fleischhack}, {Franckowiak}, {Fukazawa}, {Funk}, {Fusco}, {Galanti}, {Gammaldi}, {Gargano}, {Garrappa}, {Gasparrini}, {Giacchino}, {Giglietto}, {Giordano}, {Giroletti}, {Glanzman}, {Green}, {Grenier}, {Grondin}, {Guillemot}, {Guiriec}, {Gustafsson}, {Harding}, {Hays}, {Hewitt}, {Horan}, {Hou}, {J{\'o}hannesson}, {Karwin}, {Kayanoki}, {Kerr}, {Kuss}, {Landriu}, {Larsson}, {Latronico}, {Lemoine-Goumard}, {Li}, {Liodakis}, {Longo}, {Loparco}, {Lott}, {Lubrano}, {Maldera}, {Malyshev}, {Manfreda}, {Mart{\'\i}-Devesa},
  {Mazziotta}, {Mereu}, {Meyer}, {Michelson}, {Mirabal}, {Mitthumsiri}, {Mizuno}, {Moiseev}, {Monzani}, {Morselli}, {Moskalenko}, {Negro}, {Nuss}, {Omodei}, {Orienti}, {Orlando}, {Paneque}, {Pei}, {Perkins}, {Persic}, {Pesce-Rollins}, {Petrosian}, {Pillera}, {Poon}, {Porter}, {Principe}, {Rain{\`o}}, {Rando}, {Rani}, {Razzano}, {Razzaque}, {Reimer}, {Reimer}, {Reposeur}, {S{\'a}nchez-Conde}, {Saz Parkinson}, {Scotton}, {Serini}, {Sgr{\`o}}, {Siskind}, {Smith}, {Spandre}, {Spinelli}, {Sueoka}, {Suson}, {Tajima}, {Tak}, {Thayer}, {Thompson}, {Torres}, {Troja}, {Valverde}, {Wood}, \& {Zaharijas}}]{FermiDR3_2022}
{Abdollahi}, S., {Acero}, F., {Baldini}, L., {et~al.} 2022, \apjs, 260, 53

\bibitem[{{Anderson} {et~al.}(2014){Anderson}, {Bania}, {Balser}, {Cunningham}, {Wenger}, {Johnstone}, \& {Armentrout}}]{Anderson2014}
{Anderson}, L.~D., {Bania}, T.~M., {Balser}, D.~S., {et~al.} 2014, \apjs, 212, 1

\bibitem[{{Anderson} {et~al.}(2012){Anderson}, {Bania}, {Balser}, \& {Rood}}]{Anderson2012}
{Anderson}, L.~D., {Bania}, T.~M., {Balser}, D.~S., \& {Rood}, R.~T. 2012, \apj, 754, 62

\bibitem[{{Ballet} {et~al.}(2023){Ballet}, {Bruel}, {Burnett}, {Lott}, \& {The Fermi-LAT collaboration}}]{FermiDR4_2023}
{Ballet}, J., {Bruel}, P., {Burnett}, T.~H., {Lott}, B., \& {The Fermi-LAT collaboration}. 2023, arXiv e-prints, arXiv:2307.12546

\bibitem[{{Benaglia} {et~al.}(2021){Benaglia}, {del Palacio}, {Hales}, \& {Colazo}}]{Benaglia2021}
{Benaglia}, P., {del Palacio}, S., {Hales}, C., \& {Colazo}, M.~E. 2021, \mnras, 503, 2514

\bibitem[{{Benaglia} {et~al.}(2010){Benaglia}, {Romero}, {Mart{\'\i}}, {Peri}, \& {Araudo}}]{Benaglia2010}
{Benaglia}, P., {Romero}, G.~E., {Mart{\'\i}}, J., {Peri}, C.~S., \& {Araudo}, A.~T. 2010, \aap, 517, L10

\bibitem[{{Binder} {et~al.}(2019){Binder}, {Behr}, \& {Povich}}]{Binder2019}
{Binder}, B.~A., {Behr}, P., \& {Povich}, M.~S. 2019, \aj, 157, 176

\bibitem[{{Blaauw}(1961)}]{Blaauw1961}
{Blaauw}, A. 1961, \bain, 15, 265

\bibitem[{{Bodensteiner} {et~al.}(2018){Bodensteiner}, {Baade}, {Greiner}, \& {Langer}}]{Bodensteiner2018}
{Bodensteiner}, J., {Baade}, D., {Greiner}, J., \& {Langer}, N. 2018, \aap, 618, A110

\bibitem[{{Bordas}(2023)}]{Bordas2023}
{Bordas}, P. 2023, in 7th Heidelberg International Symposium on High-Energy Gamma-Ray Astronomy (Gamma2022), 133

\bibitem[{{Bosch-Ramon} {et~al.}(2012){Bosch-Ramon}, {Barkov}, {Khangulyan}, \& {Perucho}}]{Valenti2012}
{Bosch-Ramon}, V., {Barkov}, M.~V., {Khangulyan}, D., \& {Perucho}, M. 2012, \aap, 544, A59

\bibitem[{{Boubert} \& {Evans}(2018)}]{Boubert2018}
{Boubert}, D. \& {Evans}, N.~W. 2018, \mnras, 477, 5261

\bibitem[{{Bourke} {et~al.}(2005){Bourke}, {Hyland}, \& {Robinson}}]{Bourke2005}
{Bourke}, T.~L., {Hyland}, A.~R., \& {Robinson}, G. 2005, \apj, 625, 883

\bibitem[{{Brown} \& {Bomans}(2005)}]{BrownBomans2005}
{Brown}, D. \& {Bomans}, D.~J. 2005, \aap, 439, 183

\bibitem[{{Carretero-Castrillo} {et~al.}(2023){Carretero-Castrillo}, {Rib{\'o}}, \& {Paredes}}]{MCC2023}
{Carretero-Castrillo}, M., {Rib{\'o}}, M., \& {Paredes}, J.~M. 2023, \aap, 679, A109

\bibitem[{{Chick} {et~al.}(2020){Chick}, {Kobulnicky}, {Schurhammer}, {Andrews}, {Povich}, {Buser}, {Dixon}, {Lindman}, {Munari}, {Olivier}, {Sorber}, \& {Wernke}}]{Chick2020}
{Chick}, W.~T., {Kobulnicky}, H.~A., {Schurhammer}, D.~P., {et~al.} 2020, \apjs, 251, 29

\bibitem[{{Comer{\'o}n} \& {Pasquali}(2007)}]{Comeron2007}
{Comer{\'o}n}, F. \& {Pasquali}, A. 2007, \aap, 467, L23

\bibitem[{{Condon} {et~al.}(1998){Condon}, {Cotton}, {Greisen}, {Yin}, {Perley}, {Taylor}, \& {Broderick}}]{NVSS1998}
{Condon}, J.~J., {Cotton}, W.~D., {Greisen}, E.~W., {et~al.} 1998, \aj, 115, 1693

\bibitem[{{de Bruijne} \& {Eilers}(2012)}]{Bruijne2012}
{de Bruijne}, J.~H.~J. \& {Eilers}, A.~C. 2012, \aap, 546, A61

\bibitem[{{del Palacio} {et~al.}(2018){del Palacio}, {Bosch-Ramon}, {M{\"u}ller}, \& {Romero}}]{delpalacio2018}
{del Palacio}, S., {Bosch-Ramon}, V., {M{\"u}ller}, A.~L., \& {Romero}, G.~E. 2018, \aap, 617, A13

\bibitem[{{del Valle} \& {Romero}(2012)}]{Delvalle2012}
{del Valle}, M.~V. \& {Romero}, G.~E. 2012, \aap, 543, A56

\bibitem[{{Dubus}(2006)}]{Dubus2006}
{Dubus}, G. 2006, \aap, 456, 801

\bibitem[{{Dubus}(2013)}]{Dubus2013}
{Dubus}, G. 2013, \aapr, 21, 64

\bibitem[{{Duchesne} {et~al.}(2024){Duchesne}, {Grundy}, {Heald}, {Lenc}, {Leung}, {McConnell}, {Murphy}, {Pritchard}, {Rose}, {Thomson}, {Wang}, {Wang}, \& {Whiting}}]{RACSmid2024}
{Duchesne}, S.~W., {Grundy}, J.~A., {Heald}, G.~H., {et~al.} 2024, \pasa, 41, e003

\bibitem[{{Fender} \& {Mu{\~n}oz-Darias}(2016)}]{Fender2016}
{Fender}, R. \& {Mu{\~n}oz-Darias}, T. 2016, in Lecture Notes in Physics, Berlin Springer Verlag, ed. F.~{Haardt}, V.~{Gorini}, U.~{Moschella}, A.~{Treves}, \& M.~{Colpi}, Vol. 905, 65

\bibitem[{{Gabici} {et~al.}(2007){Gabici}, {Aharonian}, \& {Blasi}}]{Gabici2007}
{Gabici}, S., {Aharonian}, F.~A., \& {Blasi}, P. 2007, \apss, 309, 365

\bibitem[{{Gull} \& {Sofia}(1979)}]{Gull1979}
{Gull}, T.~R. \& {Sofia}, S. 1979, \apj, 230, 782

\bibitem[{{H.~E.~S.~S. Collaboration} {et~al.}(2018){H.~E.~S.~S. Collaboration}, {Abdalla}, {Abramowski}, {Aharonian}, {Ait Benkhali}, {Akhperjanian}, {Andersson}, {Ang{\"u}ner}, {Arakawa}, {Arrieta}, {Aubert}, {Backes}, {Balzer}, {Barnard}, {Becherini}, {Becker Tjus}, {Berge}, {Bernhard}, {Bernl{\"o}hr}, {Blackwell}, {B{\"o}ttcher}, {Boisson}, {Bolmont}, {Bordas}, {Bregeon}, {Brun}, {Brun}, {Bryan}, {B{\"u}chele}, {Bulik}, {Capasso}, {Carr}, {Casanova}, {Cerruti}, {Chakraborty}, {Chalme-Calvet}, {Chaves}, {Chen}, {Chevalier}, {Chr{\'e}tien}, {Coffaro}, {Colafrancesco}, {Cologna}, {Condon}, {Conrad}, {Cui}, {Davids}, {Decock}, {Degrange}, {Deil}, {Devin}, {deWilt}, {Dirson}, {Djannati-Ata{\"\i}}, {Domainko}, {Donath}, {Drury}, {Dutson}, {Dyks}, {Edwards}, {Egberts}, {Eger}, {Ernenwein}, {Eschbach}, {Farnier}, {Fegan}, {Fernandes}, {Fiasson}, {Fontaine}, {F{\"o}rster}, {Funk}, {F{\"u}{\ss}ling}, {Gabici}, {Gajdus}, {Gallant}, {Garrigoux}, {Giavitto}, {Giebels}, {Glicenstein}, {Gottschall}, {Goyal}, {Grondin},
  {Hahn}, {Haupt}, {Hawkes}, {Heinzelmann}, {Henri}, {Hermann}, {Hervet}, {Hinton}, {Hofmann}, {Hoischen}, {Holler}, {Horns}, {Ivascenko}, {Iwasaki}, {Jacholkowska}, {Jamrozy}, {Janiak}, {Jankowsky}, {Jankowsky}, {Jingo}, {Jogler}, {Jouvin}, {Jung-Richardt}, {Kastendieck}, {Katarzy{\'n}ski}, {Katsuragawa}, {Katz}, {Kerszberg}, {Khangulyan}, {Kh{\'e}lifi}, {Kieffer}, {King}, {Klepser}, {Klochkov}, {Klu{\'z}niak}, {Kolitzus}, {Komin}, {Kosack}, {Krakau}, {Kraus}, {Kr{\"u}ger}, {Laffon}, {Lamanna}, {Lau}, {Lees}, {Lefaucheur}, {Lefranc}, {Lemi{\`e}re}, {Lemoine-Goumard}, {Lenain}, {Leser}, {Lohse}, {Lorentz}, {Liu}, {L{\'o}pez-Coto}, {Lypova}, {Marandon}, {Marcowith}, {Mariaud}, {Marx}, {Maurin}, {Maxted}, {Mayer}, {Meintjes}, {Meyer}, {Mitchell}, {Moderski}, {Mohamed}, {Mohrmann}, {Mor{\r{a}}}, {Moulin}, {Murach}, {Nakashima}, {de Naurois}, {Niederwanger}, {Niemiec}, {Oakes}, {O'Brien}, {Odaka}, {{\"O}ttl}, {Ohm}, {Ostrowski}, {Oya}, {Padovani}, {Panter}, {Parsons}, {Pekeur}, {Pelletier}, {Perennes},
  {Petrucci}, {Peyaud}, {Piel}, {Pita}, {Poon}, {Prokhorov}, {Prokoph}, {P{\"u}hlhofer}, {Punch}, {Quirrenbach}, {Raab}, {Reimer}, {Reimer}, {Renaud}, {de los Reyes}, {Richter}, {Rieger}, {Romoli}, {Rowell}, {Rudak}, {Rulten}, {Sahakian}, {Saito}, {Salek}, {Sanchez}, {Santangelo}, {Sasaki}, {Schlickeiser}, {Sch{\"u}ssler}, {Schulz}, {Schwanke}, {Schwemmer}, {Seglar-Arroyo}, {Settimo}, {Seyffert}, {Shafi}, {Shilon}, {Simoni}, {Sol}, {Spanier}, {Spengler}, {Spies}, {Stawarz}, {Steenkamp}, {Stegmann}, {Stycz}, {Sushch}, {Takahashi}, {Tavernet}, {Tavernier}, {Taylor}, {Terrier}, {Tibaldo}, {Tiziani}, {Tluczykont}, {Trichard}, {Tsuji}, {Tuffs}, {Uchiyama}, {van der Walt}, {van Eldik}, {van Rensburg}, {van Soelen}, {Vasileiadis}, {Veh}, {Venter}, {Viana}, {Vincent}, {Vink}, {Voisin}, {V{\"o}lk}, {Vuillaume}, {Wadiasingh}, {Wagner}, {Wagner}, {Wagner}, {White}, {Wierzcholska}, {Willmann}, {W{\"o}rnlein}, {Wouters}, {Yang}, {Zabalza}, {Zaborov}, {Zacharias}, {Zanin}, {Zdziarski}, {Zech}, {Zefi}, {Ziegler}, \&
  {{\.Z}ywucka}}]{HESS2018}
{H.~E.~S.~S. Collaboration}, {Abdalla}, H., {Abramowski}, A., {et~al.} 2018, \aap, 612, A12

\bibitem[{{Hale} {et~al.}(2021){Hale}, {McConnell}, {Thomson}, {Lenc}, {Heald}, {Hotan}, {Leung}, {Moss}, {Murphy}, {Pritchard}, {Sadler}, {Stewart}, \& {Whiting}}]{RACSlow2021}
{Hale}, C.~L., {McConnell}, D., {Thomson}, A.~J.~M., {et~al.} 2021, \pasa, 38, e058

\bibitem[{{Hartley} {et~al.}(1986){Hartley}, {Manchester}, {Smith}, {Tritton}, \& {Goss}}]{Hartley1986}
{Hartley}, M., {Manchester}, R.~N., {Smith}, R.~M., {Tritton}, S.~B., \& {Goss}, W.~M. 1986, \aaps, 63, 27

\bibitem[{{Hoogerwerf} {et~al.}(2001){Hoogerwerf}, {de Bruijne}, \& {de Zeeuw}}]{Hoogerwerf2001}
{Hoogerwerf}, R., {de Bruijne}, J.~H.~J., \& {de Zeeuw}, P.~T. 2001, \aap, 365, 49

\bibitem[{{Jayasinghe} {et~al.}(2019){Jayasinghe}, {Dixon}, {Povich}, {Binder}, {Velasco}, {Lepore}, {Xu}, {Offner}, {Kobulnicky}, {Anderson}, {Kendrew}, \& {Simpson}}]{Jayasinghe2019}
{Jayasinghe}, T., {Dixon}, D., {Povich}, M.~S., {et~al.} 2019, \mnras, 488, 1141

\bibitem[{{Joye} \& {Mandel}(2003)}]{ds92003}
{Joye}, W.~A. \& {Mandel}, E. 2003, in Astronomical Society of the Pacific Conference Series, Vol. 295, Astronomical Data Analysis Software and Systems XII, ed. H.~E. {Payne}, R.~I. {Jedrzejewski}, \& R.~N. {Hook}, 489

\bibitem[{{Kharchenko} {et~al.}(2007){Kharchenko}, {Scholz}, {Piskunov}, {R{\"o}ser}, \& {Schilbach}}]{Kharchenko2007}
{Kharchenko}, N.~V., {Scholz}, R.~D., {Piskunov}, A.~E., {R{\"o}ser}, S., \& {Schilbach}, E. 2007, Astronomische Nachrichten, 328, 889

\bibitem[{{Kobulnicky} \& {Chick}(2022)}]{Kobulnicky2022}
{Kobulnicky}, H.~A. \& {Chick}, W.~T. 2022, \aj, 164, 86

\bibitem[{{Kobulnicky} {et~al.}(2016){Kobulnicky}, {Chick}, {Schurhammer}, {Andrews}, {Povich}, {Munari}, {Olivier}, {Sorber}, {Wernke}, {Dale}, \& {Dixon}}]{Kobulnicky2016}
{Kobulnicky}, H.~A., {Chick}, W.~T., {Schurhammer}, D.~P., {et~al.} 2016, \apjs, 227, 18

\bibitem[{{Kobulnicky} {et~al.}(2017){Kobulnicky}, {Schurhammer}, {Baldwin}, {Chick}, {Dixon}, {Lee}, \& {Povich}}]{Kobulnicky2017}
{Kobulnicky}, H.~A., {Schurhammer}, D.~P., {Baldwin}, D.~J., {et~al.} 2017, \aj, 154, 201

\bibitem[{{Kulkarni} \& {Heiles}(1988)}]{Kulkarni1988}
{Kulkarni}, S.~R. \& {Heiles}, C. 1988, in Galactic and Extragalactic Radio Astronomy, ed. K.~I. {Kellermann} \& G.~L. {Verschuur}, 95--153

\bibitem[{{Lacy} {et~al.}(2020){Lacy}, {Baum}, {Chandler}, {Chatterjee}, {Clarke}, {Deustua}, {English}, {Farnes}, {Gaensler}, {Gugliucci}, {Hallinan}, {Kent}, {Kimball}, {Law}, {Lazio}, {Marvil}, {Mao}, {Medlin}, {Mooley}, {Murphy}, {Myers}, {Osten}, {Richards}, {Rosolowsky}, {Rudnick}, {Schinzel}, {Sivakoff}, {Sjouwerman}, {Taylor}, {White}, {Wrobel}, {Andernach}, {Beasley}, {Berger}, {Bhatnager}, {Birkinshaw}, {Bower}, {Brandt}, {Brown}, {Burke-Spolaor}, {Butler}, {Comerford}, {Demorest}, {Fu}, {Giacintucci}, {Golap}, {G{\"u}th}, {Hales}, {Hiriart}, {Hodge}, {Horesh}, {Ivezi{\'c}}, {Jarvis}, {Kamble}, {Kassim}, {Liu}, {Loinard}, {Lyons}, {Masters}, {Mezcua}, {Moellenbrock}, {Mroczkowski}, {Nyland}, {O'Dea}, {O'Sullivan}, {Peters}, {Radford}, {Rao}, {Robnett}, {Salcido}, {Shen}, {Sobotka}, {Witz}, {Vaccari}, {van Weeren}, {Vargas}, {Williams}, \& {Yoon}}]{VLASS2020}
{Lacy}, M., {Baum}, S.~A., {Chandler}, C.~J., {et~al.} 2020, \pasp, 132, 035001

\bibitem[{{Landau} \& {Lifshitz}(1959)}]{Landau1959}
{Landau}, L.~D. \& {Lifshitz}, E.~M. 1959, {Fluid mechanics}

\bibitem[{{Launhardt} {et~al.}(2010){Launhardt}, {Nutter}, {Ward-Thompson}, {Bourke}, {Henning}, {Khanzadyan}, {Schmalzl}, {Wolf}, \& {Zylka}}]{Launhardt2010}
{Launhardt}, R., {Nutter}, D., {Ward-Thompson}, D., {et~al.} 2010, \apjs, 188, 139

\bibitem[{{Mackey}(2023)}]{Mackye2023}
{Mackey}, J. 2023, in Winds of Stars and Exoplanets, ed. A.~A. {Vidotto}, L.~{Fossati}, \& J.~S. {Vink}, Vol. 370, 205--216

\bibitem[{{Ma{\'\i}z Apell{\'a}niz} {et~al.}(2018){Ma{\'\i}z Apell{\'a}niz}, {Pantaleoni Gonz{\'a}lez}, {Barb{\'a}}, {Sim{\'o}n-D{\'\i}az}, {Negueruela}, {Lennon}, {Sota}, \& {Trigueros P{\'a}ez}}]{MA2018}
{Ma{\'\i}z Apell{\'a}niz}, J., {Pantaleoni Gonz{\'a}lez}, M., {Barb{\'a}}, R.~H., {et~al.} 2018, \aap, 616, A149

\bibitem[{{Ma{\'\i}z Apell{\'a}niz} {et~al.}(2013){Ma{\'\i}z Apell{\'a}niz}, {Sota}, {Morrell}, {Barb{\'a}}, {Walborn}, {Alfaro}, {Gamen}, {Arias}, \& {Gallego Calvente}}]{GOSC}
{Ma{\'\i}z Apell{\'a}niz}, J., {Sota}, A., {Morrell}, N.~I., {et~al.} 2013, in Massive Stars: From alpha to Omega, 198

\bibitem[{{Marcote} {et~al.}(2018){Marcote}, {Rib{\'o}}, {Paredes}, {Mao}, \& {Edwards}}]{Marcote2018}
{Marcote}, B., {Rib{\'o}}, M., {Paredes}, J.~M., {Mao}, M.~Y., \& {Edwards}, P.~G. 2018, \aap, 619, A26

\bibitem[{{Martinez} {et~al.}(2023){Martinez}, {del Palacio}, \& {Bosch-Ramon}}]{Martinez2023}
{Martinez}, J.~R., {del Palacio}, S., \& {Bosch-Ramon}, V. 2023, \aap, 680, A99

\bibitem[{{Martins} {et~al.}(2005){Martins}, {Schaerer}, \& {Hillier}}]{Martins2005}
{Martins}, F., {Schaerer}, D., \& {Hillier}, D.~J. 2005, \aap, 436, 1049

\bibitem[{{McConnell} {et~al.}(2020){McConnell}, {Hale}, {Lenc}, {Banfield}, {Heald}, {Hotan}, {Leung}, {Moss}, {Murphy}, {O'Brien}, {Pritchard}, {Raja}, {Sadler}, {Stewart}, {Thomson}, {Whiting}, {Allison}, {Amy}, {Anderson}, {Ball}, {Bannister}, {Bell}, {Bock}, {Bolton}, {Bunton}, {Chippendale}, {Collier}, {Cooray}, {Cornwell}, {Diamond}, {Edwards}, {Gupta}, {Hayman}, {Heywood}, {Jackson}, {Koribalski}, {Lee-Waddell}, {McClure-Griffiths}, {Ng}, {Norris}, {Phillips}, {Reynolds}, {Roxby}, {Schinckel}, {Shields}, {Tremblay}, {Tzioumis}, {Voronkov}, \& {Westmeier}}]{RACS2020}
{McConnell}, D., {Hale}, C.~L., {Lenc}, E., {et~al.} 2020, \pasa, 37, e048

\bibitem[{{M{\'e}ndez-Delgado} {et~al.}(2022){M{\'e}ndez-Delgado}, {Amayo}, {Arellano-C{\'o}rdova}, {Esteban}, {Garc{\'\i}a-Rojas}, {Carigi}, \& {Delgado-Inglada}}]{MendezDelgado2022}
{M{\'e}ndez-Delgado}, J.~E., {Amayo}, A., {Arellano-C{\'o}rdova}, K.~Z., {et~al.} 2022, \mnras, 510, 4436

\bibitem[{{Miller-Jones} {et~al.}(2018){Miller-Jones}, {Deller}, {Shannon}, {Dodson}, {Mold{\'o}n}, {Rib{\'o}}, {Dubus}, {Johnston}, {Paredes}, {Ransom}, \& {Tomsick}}]{Miller-Jones2018}
{Miller-Jones}, J.~C.~A., {Deller}, A.~T., {Shannon}, R.~M., {et~al.} 2018, \mnras, 479, 4849

\bibitem[{{Mirabel} \& {Rodr{\'\i}guez}(1999)}]{Mirabel1999}
{Mirabel}, I.~F. \& {Rodr{\'\i}guez}, L.~F. 1999, \araa, 37, 409

\bibitem[{{Miroshnichenko} {et~al.}(2000){Miroshnichenko}, {Chentsov}, \& {Klochkova}}]{Miroshnichenko2000}
{Miroshnichenko}, A.~S., {Chentsov}, E.~L., \& {Klochkova}, V.~G. 2000, \aaps, 144, 379

\bibitem[{{Moutzouri} {et~al.}(2022){Moutzouri}, {Mackey}, {Carrasco-Gonz{\'a}lez}, {Gong}, {Brose}, {Zargaryan}, {Toal{\'a}}, {Menten}, {Gvaramadze}, \& {Rugel}}]{Moutzouri2022}
{Moutzouri}, M., {Mackey}, J., {Carrasco-Gonz{\'a}lez}, C., {et~al.} 2022, \aap, 663, A80

\bibitem[{{Neiner} {et~al.}(2011){Neiner}, {de Batz}, {Cochard}, {Floquet}, {Mekkas}, \& {Desnoux}}]{BeSS}
{Neiner}, C., {de Batz}, B., {Cochard}, F., {et~al.} 2011, \aj, 142, 149

\bibitem[{{Noriega-Crespo} {et~al.}(1997){Noriega-Crespo}, {van Buren}, \& {Dgani}}]{NoriegaCrespo1997}
{Noriega-Crespo}, A., {van Buren}, D., \& {Dgani}, R. 1997, \aj, 113, 780

\bibitem[{{Ogrodnikoff}(1932)}]{Ogrodnikoff1932}
{Ogrodnikoff}, K. 1932, \zap, 4, 190

\bibitem[{{Omar} {et~al.}(2002){Omar}, {Chengalur}, \& {Anish Roshi}}]{Omar2002}
{Omar}, A., {Chengalur}, J.~N., \& {Anish Roshi}, D. 2002, \aap, 395, 227

\bibitem[{{Oort}(1927)}]{Oort1927}
{Oort}, J.~H. 1927, \bain, 3, 275

\bibitem[{{Patra} \& {Roy}(2024)}]{PatraRoy2024}
{Patra}, N.~N. \& {Roy}, N. 2024, \mnras, 529, 4037

\bibitem[{{Peri} {et~al.}(2012){Peri}, {Benaglia}, {Brookes}, {Stevens}, \& {Isequilla}}]{Peri2012}
{Peri}, C.~S., {Benaglia}, P., {Brookes}, D.~P., {Stevens}, I.~R., \& {Isequilla}, N.~L. 2012, \aap, 538, A108

\bibitem[{{Peri} {et~al.}(2015){Peri}, {Benaglia}, \& {Isequilla}}]{Peri2015}
{Peri}, C.~S., {Benaglia}, P., \& {Isequilla}, N.~L. 2015, \aap, 578, A45

\bibitem[{{Povich} {et~al.}(2008){Povich}, {Benjamin}, {Whitney}, {Babler}, {Indebetouw}, {Meade}, \& {Churchwell}}]{Povich2008}
{Povich}, M.~S., {Benjamin}, R.~A., {Whitney}, B.~A., {et~al.} 2008, \apj, 689, 242

\bibitem[{{Prinja} {et~al.}(1990){Prinja}, {Barlow}, \& {Howarth}}]{Prinja1990}
{Prinja}, R.~K., {Barlow}, M.~J., \& {Howarth}, I.~D. 1990, \apj, 361, 607

\bibitem[{{Reid} {et~al.}(2019){Reid}, {Menten}, {Brunthaler}, {Zheng}, {Dame}, {Xu}, {Li}, {Sakai}, {Wu}, {Immer}, {Zhang}, {Sanna}, {Moscadelli}, {Rygl}, {Bartkiewicz}, {Hu}, {Quiroga-Nu{\~n}ez}, \& {van Langevelde}}]{Reid2019}
{Reid}, M.~J., {Menten}, K.~M., {Brunthaler}, A., {et~al.} 2019, \apj, 885, 131

\bibitem[{{Rib{\'o}} {et~al.}(2002){Rib{\'o}}, {Paredes}, {Romero}, {Benaglia}, {Mart{\'\i}}, {Fors}, \& {Garc{\'\i}a-S{\'a}nchez}}]{Ribo2002}
{Rib{\'o}}, M., {Paredes}, J.~M., {Romero}, G.~E., {et~al.} 2002, \aap, 384, 954

\bibitem[{{Rivinius} {et~al.}(2013){Rivinius}, {Carciofi}, \& {Martayan}}]{Rivinius2013}
{Rivinius}, T., {Carciofi}, A.~C., \& {Martayan}, C. 2013, \aapr, 21, 69

\bibitem[{{Scheffler} \& {Elsässer}(1987)}]{SchefflerElsasser1987}
{Scheffler}, H. \& {Elsässer}, H. 1987, Physics of the Galaxy and Interstellar Matter (Springer-Verlag)

\bibitem[{{Schinckel} {et~al.}(2012){Schinckel}, {Bunton}, {Cornwell}, {Feain}, \& {Hay}}]{ASKAP2012}
{Schinckel}, A.~E., {Bunton}, J.~D., {Cornwell}, T.~J., {Feain}, I., \& {Hay}, S.~G. 2012, in Society of Photo-Optical Instrumentation Engineers (SPIE) Conference Series, Vol. 8444, Ground-based and Airborne Telescopes IV, ed. L.~M. {Stepp}, R.~{Gilmozzi}, \& H.~J. {Hall}, 84442A

\bibitem[{{Simpson} {et~al.}(2012){Simpson}, {Povich}, {Kendrew}, {Lintott}, {Bressert}, {Arvidsson}, {Cyganowski}, {Maddison}, {Schawinski}, {Sherman}, {Smith}, \& {Wolf-Chase}}]{Simpson2012}
{Simpson}, R.~J., {Povich}, M.~S., {Kendrew}, S., {et~al.} 2012, \mnras, 424, 2442

\bibitem[{{Stone}(1979)}]{Stone1979}
{Stone}, R.~C. 1979, \apj, 232, 520

\bibitem[{Swarup {et~al.}(1991)Swarup, Ananthakrishnan, Kapahi, Rao, Subrahmanya, \& Kulkarni}]{GMRT1991}
Swarup, G., Ananthakrishnan, S., Kapahi, V.~K., {et~al.} 1991, Current Science, 60, 95

\bibitem[{{Tetzlaff} {et~al.}(2011){Tetzlaff}, {Neuh{\"a}user}, \& {Hohle}}]{Tetzlaff2011}
{Tetzlaff}, N., {Neuh{\"a}user}, R., \& {Hohle}, M.~M. 2011, \mnras, 410, 190

\bibitem[{{van den Eijnden} {et~al.}(2022{\natexlab{a}}){van den Eijnden}, {Heywood}, {Fender}, {Mohamed}, {Sivakoff}, {Saikia}, {Russell}, {Motta}, {Miller-Jones}, \& {Woudt}}]{VandenEijnden2022a}
{van den Eijnden}, J., {Heywood}, I., {Fender}, R., {et~al.} 2022{\natexlab{a}}, \mnras, 510, 515

\bibitem[{{van den Eijnden} {et~al.}(2022{\natexlab{b}}){van den Eijnden}, {Saikia}, \& {Mohamed}}]{VandenEijnden2022b}
{van den Eijnden}, J., {Saikia}, P., \& {Mohamed}, S. 2022{\natexlab{b}}, \mnras, 512, 5374

\bibitem[{{Vink} \& {Sander}(2021)}]{Vink2021}
{Vink}, J.~S. \& {Sander}, A. A.~C. 2021, \mnras, 504, 2051

\bibitem[{{Wang} {et~al.}(2023){Wang}, {Wu}, {Jiang}, \& {Zhang}}]{Wang2023}
{Wang}, M., {Wu}, J., {Jiang}, B., \& {Zhang}, Y. 2023, \apjs, 267, 39

\bibitem[{{Wilkin}(1996)}]{Wilkin1996}
{Wilkin}, F.~P. 1996, \apjl, 459, L31

\bibitem[{{Williams} {et~al.}(2011){Williams}, {Gies}, {Hillwig}, {McSwain}, \& {Huang}}]{Williams2011}
{Williams}, S.~J., {Gies}, D.~R., {Hillwig}, T.~C., {McSwain}, M.~V., \& {Huang}, W. 2011, \aj, 142, 146

\bibitem[{{Wright} {et~al.}(2010){Wright}, {Eisenhardt}, {Mainzer}, {Ressler}, {Cutri}, {Jarrett}, {Kirkpatrick}, {Padgett}, {McMillan}, {Skrutskie}, {Stanford}, {Cohen}, {Walker}, {Mather}, {Leisawitz}, {Gautier}, {McLean}, {Benford}, {Lonsdale}, {Blain}, {Mendez}, {Irace}, {Duval}, {Liu}, {Royer}, {Heinrichsen}, {Howard}, {Shannon}, {Kendall}, {Walsh}, {Larsen}, {Cardon}, {Schick}, {Schwalm}, {Abid}, {Fabinsky}, {Naes}, \& {Tsai}}]{WISE2010}
{Wright}, E.~L., {Eisenhardt}, P. R.~M., {Mainzer}, A.~K., {et~al.} 2010, \aj, 140, 1868

\end{thebibliography}

\begin{appendix}

\onecolumn
\section{WISE RGB images for new bow shock and bubbles} \label{Sec:RGBs_New_App}

\begin{figure*}[h!!!]
\centerline{\includegraphics[width=0.88\hsize]{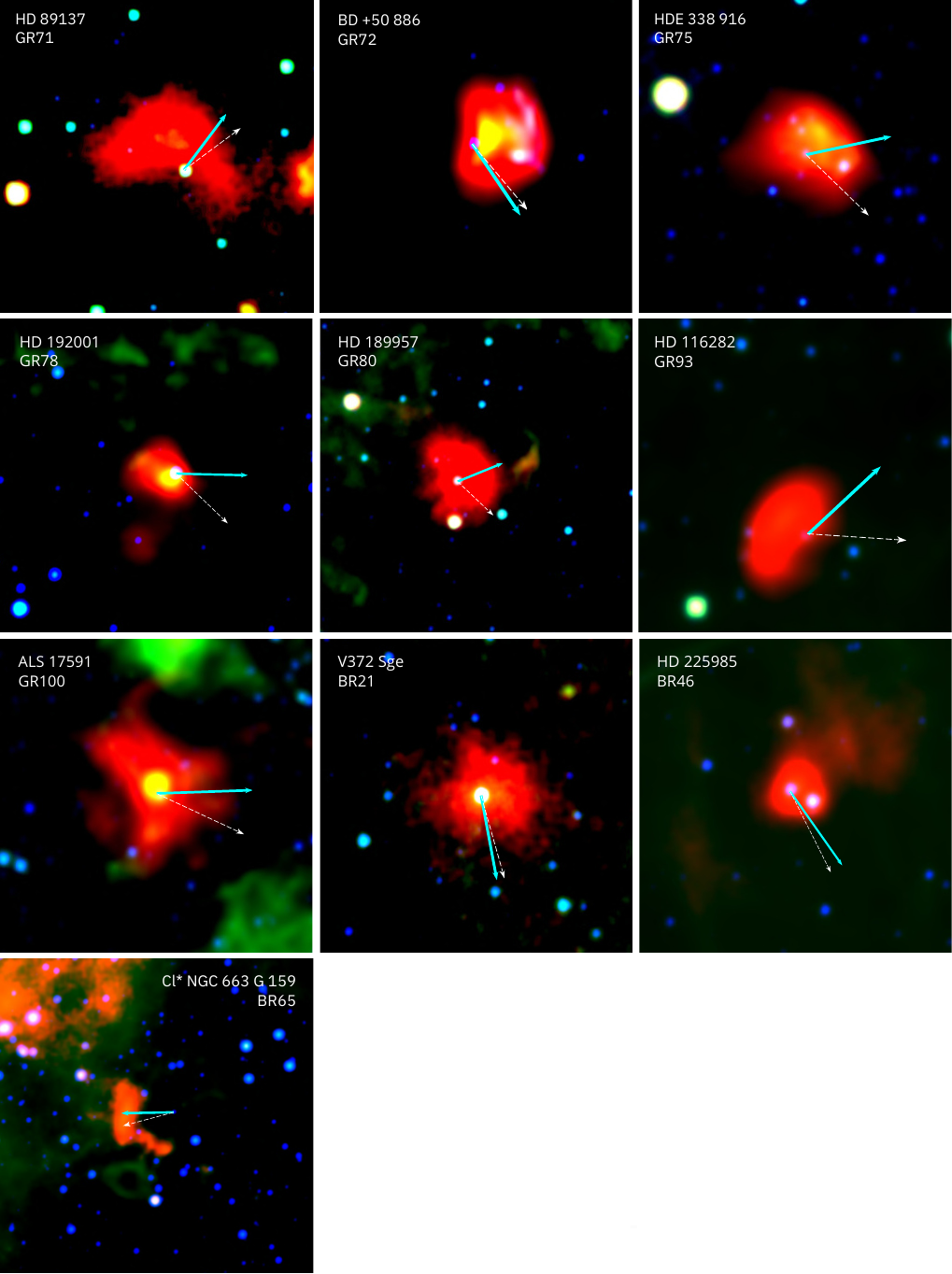}}
\caption{WISE RGB images in W4+W3+W2 for the new bow shock and bubble candidates presented in Table~\ref{Tab:NewBowShocks} in equatorial coordinates with North up and East to the left. Dashed white arrows indicate the directions of the original proper motions from \textit{Gaia}~DR3, and solid cyan arrows indicate the directions of the proper motions corrected for the ISM motion caused by Galactic rotation. Each field has a different size, but the arrows are fixed to 2\arcmin. Each panel contains the name of the star as provided in the GOSC or the BeSS catalogs, and the runaway star id. The RGB images of the first three sources from Table~\ref{Tab:NewBowShocks} are presented in Fig.~\ref{Fig:RGBs_New_MainText}.}
\label{Fig:RGBs_New_App}
\end{figure*}

\FloatBarrier


\section{Bow shock direction differences} \label{Sec:App_DirectionDifferences}

As can be seen in some of the bow shocks shown in Figs.~\ref{Fig:RGBs_New_MainText}, \ref{Fig:RGBs_New_App}, and~\ref{Fig:RGBs_Known}, the corrected proper motion direction of the runaway star is not well aligned with the bow shock direction. We tabulate and comment on these differences here.\\

For each runaway with an associated bow shock, either new or known, we list in Table~\ref{Tab:AngularDifferences} the position angle, $PA$, computed from north to east of the corrected proper motion of the runaway star, as well as its corresponding uncertainty $\Delta{PA}$, as explained in Sect.~\ref{Sec:Methodology}. We also list the bow shock angle, BS Angle, that we estimate visually from the W4 images and indicates the direction of the bow shock in the plane of the sky, together with an assumed systematic uncertainty that we call $\Delta$BS Angle of 5\degr\ at 1$\sigma$. We also list the AD, computed as the absolute value of the difference between $PA$ and BS Angle for each bow shock, together with AD Significance, which is the significance in $\sigma$ units of the AD considering the previously quoted uncertainties.\\

For comparison, we list these values both for the new and the known bow shocks identified in our samples. For AD above $3\sigma$, there are 4 new and 5 known bow shocks, while above $5\sigma$, there are 3 new and 3 known bow shocks. This indicates that some bow shock candidates identified in previous studies also exhibit misalignments between their proper motion and bow shock directions. For the new bow shocks that have AD above $5\sigma$, we comment on their possible origin as in situ bow shocks in Sect~\ref{Sec:BS_BU_new}.

\renewcommand{\arraystretch}{1.1}
\begin{table*}[h!]
\centering
\caption{Runaway star corrected proper motion direction ($PA$) and bow shock (BS) direction angles with their uncertainties, their difference and significance, for the new and known bow shocks.}
\label{Tab:AngularDifferences}
{\begin{tabular}{l@{~}c@{~~~~~}c@{~~~~~}c@{~~~~~}c@{~~~}c@{~~~}c@{~~~}l@{~~~}}
\hline\hline \vspace{-2mm}\\
Runaway id & Type & $PA$ & $\Delta{PA}$ & BS Angle & $\Delta$BS Angle & Angular difference (AD) & AD Significance\\
&      & ($\degr$)                & ($\degr$)      & ($\degr$)          & ($\degr$)                & ($\degr$)               & \\
\hline \vspace{-2mm}\\ 
GR9         & New    &  183.6   & 2.2   & 180.4 & 5.0    &  3.2     &  0.6 \\
GR34    & New    &  276.9       & 3.3   & 276.9 & 5.0    &  0.0     &  0.0 \\
GR35    & New    &  89.6        & 4.9   & 72.5  & 5.0    &  17.2        &  2.5 \\
GR71    & New    &  323.9       & 3.1   & 17.3  & 5.0    &  53.4        &  9.1 \\
GR72    & New    &  213.1       & 7.4   & 288.5 & 5.0    &  75.3        &  8.4 \\
GR75    & New    &  282.0       & 7.8   & 309.5 & 5.0    &  27.5        &  3.0 \\
GR80    & New    &  292.8       & 5.9   & 310.9 & 5.0    &  18.1        &  2.4 \\
GR93    & New    &  313.5       & 12.4  & 54.4  & 5.0    &  100.9       &  7.5 \\
GR100   & New    &  271.1       & 5.2   & 296.6 & 5.0    &  25.5        &  3.5 \\
BR65    & New    &  90.6        & 2.0   & 90.6  & 5.0    &  0.0     &  0.0 \\
GR8         & Known  &  121.7   & 5.2   & 116.7 & 5.0    &  5.0     &  0.7 \\
GR13    & Known  &  6.4     & 2.3       & 337.1 & 5.0    &  29.3        &  5.3 \\
GR16    & Known  &  356.0       & 1.9   & 355.7 & 5.0    &  0.3     &  0.1 \\
GR22    & Known  &  226.6       & 2.4   & 174.0 & 5.0    &  52.6        &  9.5 \\
GR25    & Known  &  354.4       & 2.1   & 10.3  & 5.0    &  15.8        &  2.9 \\
GR40    & Known  &  279.5       & 6.8   & 279.5 & 5.0    &  0.0     &  0.0 \\
GR42    & Known  &  47.5        & 2.3   & 38.9  & 5.0    &  8.6     &  1.6 \\
GR50    & Known  &  171.7       & 2.3   & 138.8 & 5.0    &  32.9        &  6.0 \\
GR51    & Known  &  274.3       & 9.4   & 318.7 & 5.0    &  44.4        &  4.2 \\
GR56    & Known  &  285.6       & 3.5   & 304.9 & 5.0    &  19.4        &  3.2 \\
GR65    & Known  &  24.5        & 5.1   & 11.5  & 5.0    &  13.0        &  1.8 \\
GR70    & Known  &  292.1       & 11.2  & 308.3 & 5.0    &  16.3        &  1.3 \\
GR76    & Known  &  35.1        & 8.7   & 56.7  & 5.0    &  21.6        &  2.1 \\
GR83    & Known  &  128.0       & 10.5  & 114.6 & 5.0    &  13.4        &  1.2 \\
GR92    & Known  &  279.1       & 15.5  & 287.5 & 5.0    &  8.4     &  0.5 \\
\hline
\end{tabular}}
\end{table*}

\FloatBarrier

\twocolumn
\section{Search for radio emission} \label{Sec:App_Radio}

\subsection{Radio data} \label{Sec:App_Radiodata}

Stellar bow shocks emit mainly at IR but there is some observational evidence of radio emission coming from these sources 
\citep{Benaglia2010,VandenEijnden2022a,VandenEijnden2022b,Moutzouri2022}. For the IR sources identified, we searched for observations carried out with the Very Large Array\footnote{\href{https://science.nrao.edu/facilities/vla}{https://science.nrao.edu/facilities/vla}} (VLA), the Giant Metrewave Radio Telescope (GMRT) \citep{GMRT1991}, and the Australian Square Kilometre Array Pathfinder (ASKAP) \citep{ASKAP2012}. In particular, we searched for the images provided by the NRAO VLA Sky Survey \citep[][NVSS]{NVSS1998}, the VLA Sky Survey \citep[][VLASS]{VLASS2020}, and the Rapid ASKAP Continuum Survey, in both RACS-low, and RACS-mid images \citep[][respectively]{RACSlow2021,RACSmid2024}. In addition, in the case of GMRT and VLA archival observations, we looked for projects that might have produced related papers and images. Table~\ref{Tab:Radio_Surveys} presents the main properties of the radio surveys used in this work. These include their wavelength, angular resolution and sky coverage. We note that not all our sources are visible for each of the mentioned radio interferometers due to declination limits.

\renewcommand{\arraystretch}{1.1}
\begin{table}[h]
\centering
\caption{Main properties of the radio surveys used in this work.}
\label{Tab:Radio_Surveys}
\centering
\begin{tabular}{l@{~~~}c@{~~~}c@{~~~}c@{~~~}c@{~~~}c}
\hline\hline \vspace{-3mm}\\
Survey     & Band & Wavelength   &  Resolution       & Coverage       & Ref. \\
&  &  (cm) &  (\arcsec)   &    &   \\
\hline\vspace{-3mm}\\
NVSS       &    & 21        & 45          &  $\delta\gtrsim -40\degr$   & 1 \\
VLASS      &    &  10      & 2.5         &  $\delta\gtrsim -40\degr$   & 2 \\
RACS       & low &  34     & $\sim$15    &  $\delta\lesssim +49\degr$  & 3 \\
RACS       & mid &  22     & $\sim$10    &  $\delta\lesssim +49\degr$  & 4 \\
\hline
\end{tabular}
\tablebib{(1)~\citet{NVSS1998}; (2)~\citet{VLASS2020}; (3)~\citet{RACSlow2021}; (4)~\citet{RACSmid2024}.}
\end{table}

\subsection{Search for radio counterparts} \label{Sec:App_Radiosearch}

We performed a search for radio observations for the 17 objects listed in Tables~\ref{Tab:NewBowShocks} and \ref{Tab:KnownBowShocks}, together with the \object{BS~J075339$-$2616.5} candidate. In the various radio archival data presented in \ref{Sec:App_Radiodata}, we looked for radio emission that could be superimposed in the plane of the sky, either on the runaway stars or on the bow shock and bubble candidates, and focusing in the direction of the movement of the runaway star. No discrete radio emission was detected at the specific positions of the runaway stars. We include a summary of the findings for each of these sources in Table.~\ref{Tab:RadioSearchResults}. For a few of them, the images presented reduction artifacts, precluding any findings (ill-imaged). We note that the fact that some emission was found does not indicate that it is a radio counterpart of the IR source. Out of the 18 objects, we comment on 6 cases where we found radio emission partially overlapping the IR source at some frequency (even if not significant in some cases). For the most interesting cases, we include the corresponding RACS images since they have better angular resolution than those of NVSS, and VLASS images are optimized for discrete rather than extended emission. The sources are presented in the same order as in Table~\ref{Tab:RadioSearchResults}.

\begin{table*}
\centering
\small
\caption{Results of the search for radio emission overlapping the new and the known bow shock (BS) and bubble (BU) candidates in different radio surveys.}
\label{Tab:RadioSearchResults}
\centering
{\begin{tabular}{l@{~~~}r@{~}c@{~}l@{~~~}r@{~}c@{~}l@{~~~~~~} l l l l}
\hline\hline \vspace{-2mm}\\
BS/BU       & \multicolumn{3}{c}{RA} & \multicolumn{3}{c}{DEC}                   & NVSS       & VLASS       & RACS-low      & RACS-mid \\
            & \multicolumn{3}{c}{(h m s)} & \multicolumn{3}{c}{($\degr$  $\arcmin$  $\arcsec$)}  &            &             &               &           \\
\hline 
\noalign{\smallskip}
\multicolumn{11}{c}{New BS and BU candidates associated with O-type runaway stars} \\
\noalign{\smallskip}
\hline
BS-GR9\tablefootmark{*}           & 06 & 08 & 55.822 & 15    & 42 & 18.02  & no emission      &  no emission      & no emission   & no emission   \\
BS-GR34                           & 20 & 22 & 37.757 & 41    & 40 & 29.19  & no emission      &  discrete sources & no images     & discrete sources  \\
BS-GR35                           & 07 & 53 & 38.205 & $-$26 & 14 & 02.53  & no emission      &  no emission      & no emission   & no emission   \\
BS-GR71                           & 10 & 15 & 40.071 & $-$51 & 15 & 23.99  & ---              &  ---              & no emission   & no emission   \\
BS-GR72\tablefootmark{*}          & 04 & 03 & 20.746 & 51    & 18 & 52.49  & extended source  &  ill-imaged       & ---           & ---   \\
BS-GR75\tablefootmark{*}          & 19 & 45 & 42.111 & 25    & 21 & 16.35  & no emission      &  ill-imaged       & no emission   & no emission   \\
BU-GR78                           & 20 & 11 & 01.700 & 42    & 07 & 36.33  & no emission      &  no emission      & no images     & no emission   \\
BS-GR80                           & 20 & 00 & 59.998 & 42    & 00 & 30.76  & unrelated emission  &  discrete source   & no images     & ill-imaged    \\
BS-GR93                           & 13 & 23 & 56.244 & $-$59 & 48 & 34.82  & ---      &  ---   & no emission   & no emission   \\
BU/BS-GR100                       & 15 & 13 & 55.196 & $-$59 & 07 & 51.64  & ---      &  ---   & no emission   & no emission   \\
\hline 
\noalign{\smallskip}
\multicolumn{11}{c}{New BS and BU candidates associated with Be-type runaway stars} \\
\noalign{\smallskip}
\hline
BU-BR21                           & 20 & 09 & 39.591 & 21    & 04 & 43.49  & no emission      &  no emission   & no emission   & no emission    \\
BU-BR46                           & 19 & 49 & 32.908 & 32    & 57 & 22.16  & no emission      &  no emission   & no emission   & no emission   \\
BS-BR65                           & 01 & 45 & 11.716 & 61    & 10 & 22.89  & no emission      &  no emission   & ---   & ---  \\
\hline 
\noalign{\smallskip}
\multicolumn{11}{c}{New BS candidate found in the field of BS-GR35 (no runaway star associated)} \\
\noalign{\smallskip}
\hline \vspace{-3mm} \\
BS J075339$-$2616.5\tablefootmark{*} & 07 & 53 & 39.1 & $-$26 & 16 & 31  & weak emission    &  no emission   & emission      & weak emission  \\
\hline 
\noalign{\smallskip}
\multicolumn{11}{c}{Known BS and BU candidates} \\
\noalign{\smallskip}
\hline     
BS-GR40                           & 06 & 34 & 23.562 & 02    & 32 & 02.96  & no emission      &  no emission   & no emission   & no emission  \\
BS/BU-GR51\tablefootmark{*}          & 18 & 30 & 34.785 & $-$08 & 38 & 03.72  & emission         &  no emission   & emission      & weak emission \\
BS-GR70                           & 07 & 53 & 01.006 & $-$27 & 06 & 57.70  & no emission      &  no emission   & no emission   & no emission \\
BS-GR83\tablefootmark{*}          & 17 & 15 & 22.330 & $-$38 & 12 & 46.79  & no emission      &  no images     & weak emission & no emission \\
\hline
\end{tabular}}
\tablefoot{The horizontal lines separate the objects in four groups as indicated in the table. RA and DEC coordinates are those of the runaway stars, except for \object{BS~J075339$-$2616.5} which are the BS coordinates. '---' means that the source is not visible for the observing instrument.\\
\tablefoottext{*}{These sources have their own discussion in Sect.~\ref{Sec:App_Radiosearch}.}}
\end{table*}

\subsubsection{BS-GR9 (HD 41997)}

In NVSS and RACS, we did not find emission at the bow shock position, but $\sim$~4\arcmin\ southward from it. We found that this emission is not associated with the bow shock but to the \object{Lower's Nebula} \object{Sh2-261}, which is an \ion{H}{II} region \citep[e.g.,][and references therein]{Wang2023}.

We highlight here that we found that BS-GR9 (and GR9) is contained in the uncertainty ellipse of the new {\it Fermi} source \object{4FGL J0609.1+1544} of the latest 4FGL-DR4 release \citep{FermiDR3_2022,FermiDR4_2023}. This High Energy (HE) gamma-ray source is not associated with any known source at other energies. Although this {\it Fermi} ellipse has a semimajor axis of 25\arcmin\ and a semiminor axis of 18\arcmin, an association of this new HE source with the bow shock would be very interesting, because it could provide additional evidence of nonthermal processes in these sources \citep{Benaglia2010}. In this context, we note the attempts by H.E.S.S. to search for Very High Energy (VHE) gamma-ray emission from bow shocks of runaway stars \citep{HESS2018}. Deeper radio observations than the available ones from RACS would be interesting in the search for nonthermal emission from this bow shock. Moreover, runaway stars can be found in binary systems containing compact objects (see Sect.~\ref{Sec:Introduction}), which can present nonthermal point-like radio emission and eventually HE and/or VHE gamma-ray emission. GR9 is not detected in the VLASS image with an RMS of $\sim$0.2~mJy~beam$^{-1}$. Thus, deeper observations could unveil the presence of a nonthermal radio source that could be connected to the HE gamma-ray one.

\subsubsection{BS-GR72 (BD $+$50 886)}

For the source BS-GR72 there are continuum images at 235, 325 and 610 MHz from GMRT data \citep{Omar2002} showing a rather strong and extended source closely overlapping the extent of the WISE bow shock candidate.
This radio source has been identified as the evolved \ion{H}{II} region \object{S206} (or \object{NGC~1491}), at a distance of $2.96^{+0.17}_{-0.15}$~kpc \citep{MendezDelgado2022}. At maximum angular resolution, its size is $\sim~4’$ or 740~pc. \citet{Omar2002} derived the electron temperature and emission measure of the source, and showed a spectral index (their Fig.~2) consistent with optically thick emission of such an object at the lower frequency bands, and a turnover around the largest frequency they observed, 610~MHz. The radio emission is also visible in the NVSS image. The VLASS images showed some emission at $\sim$10\arcsec\ scales superposed to the position of this bow shock, but with severe reduction artifacts. 
The runaway star GR72 is located at $2.94\pm0.3$~kpc \citep{MCC2023} in the same direction of S206. Therefore, this bow shock could be indeed an example of an in situ bow shock formed by ISM outflows from the \ion{H}{II} region (see Sect.~\ref{Sec:BS_BU_new}).

\subsubsection{BS-GR75 (HDE 338 916)}

For BS-GR75, the VLASS, RACS, and NVSS images show emission close to the bow shock position. However, it seems to come from a strong and extended source to the southeast, in the opposite direction of motion of the runaway. We then discarded that the mentioned emission can be a counterpart for our bow shock. That emission, in the case of VLASS, produces severe artifacts in the field, and through the bow shock position itself. The RACS fields are also quite noisy, so we did not include these radio images here. The RMS are 0.5~mJy~beam$^{-1}$ and 0.2~mJy~beam$^{-1}$ for RACS-low and -mid images, respectively. The mentioned strong radio emission hampers the possible detection of a weaker radio emission coming from our bow shock. Therefore, deeper radio observations, preferentially including also shorter baselines, could unveil a possible radio counterpart for BS-GR75 (see Sect.~\ref{Sec:App_Emission_mechanisms}).

\subsubsection{BS~J075339$-$2616.5 (--)} \label{Sec:App_Radio_BSJ075339-2616.5}

\begin{figure}[h!!!]
\centerline{\includegraphics[width=0.9\hsize]{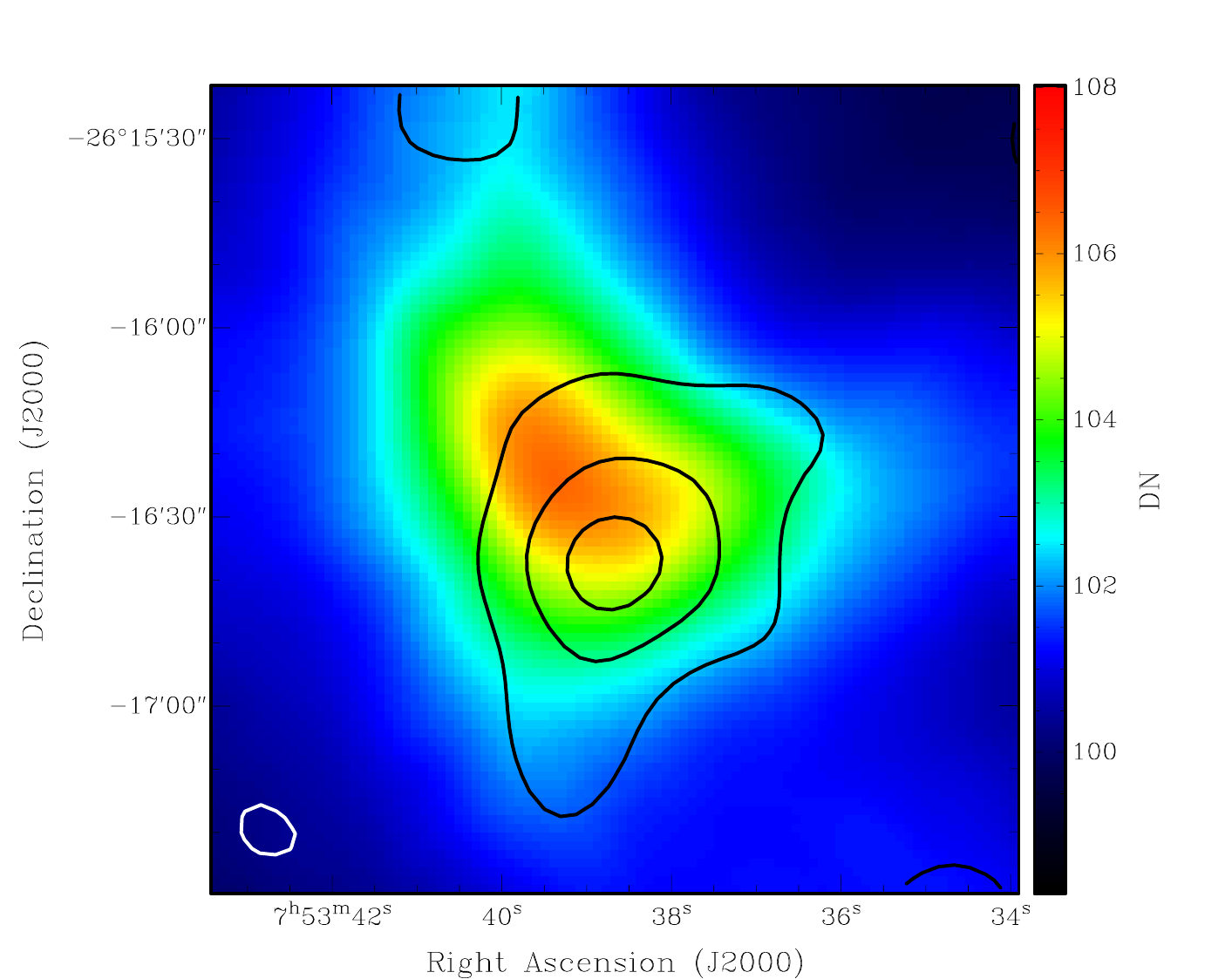}}
\centerline{\includegraphics[width=0.9\hsize]{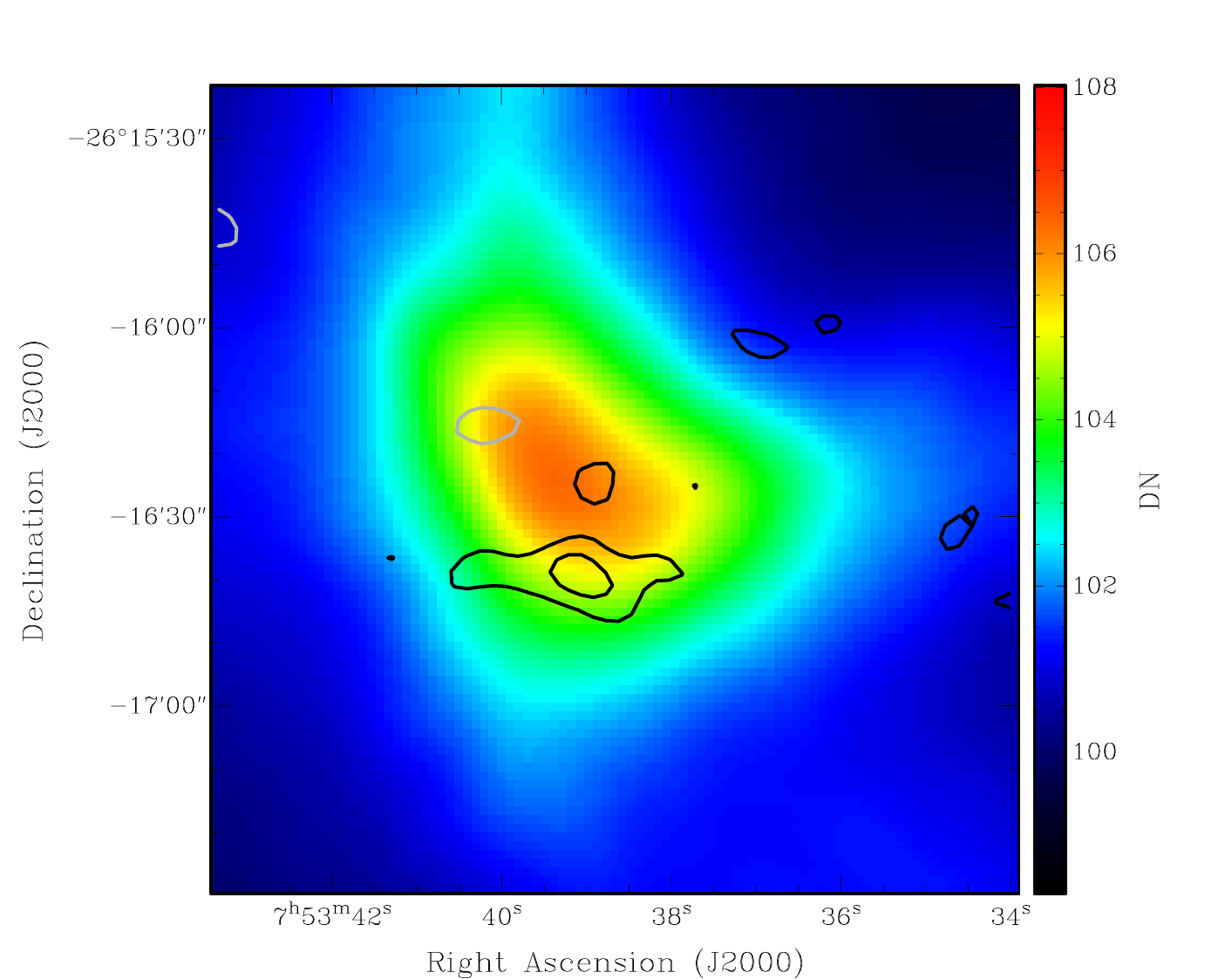}}
\caption{WISE W4 image of the bow shock candidate \object{BS~J075339$-$2616.5}. The color scale is WISE flux measured in profile-fit photometry, in units of Digital Numbers (DN; see \url{https://wise2.ipac.caltech.edu/docs/release/allsky/expsup/sec1_4c.html\#photcal}). 
{\sl Top}: RACS-low emission contours onto W4, levels at $-$0.7 (in white), 0.7, 2 and 3~mJy~beam$^{-1}$ (in black); synthesized beam of $25'' \times 25''$. {\sl Bottom}: RACS-mid emission contours onto W4, levels at $-$0.5 (in white), 0.5 and 1~mJy~beam$^{-1}$ (in black); synthesized beam of $9.4'' \times 8.3''$.}
\label{Fig:RACS_BSJ075339-2616.5}
\end{figure}

As explained in Sect.~\ref{Sec:BS_BU_new}, this bow shock candidate was found in a different way than those presented in this work. It appeared in the BS-GR35 WISE field, at $\sim$2\arcmin\ south from the GR35 runaway. The RGB image in W4+W3+W2 centered and zoomed in \object{BS~J075339$-$2616.5} is presented in Fig.~\ref{Fig:RGB_BSJ075339-2616.5}. The white cross in this figure denotes the center of the bow shock emission, whose ICRS coordinates, RA = 07h 53m 39.1s, DEC = $-$26\degr\ 16\arcmin\ 31\arcsec, give it its name. No VLASS and only weak NVSS emission were detected at the position of this source. However, it has radio emission in both RACS-low and -mid images. Figure~\ref{Fig:RACS_BSJ075339-2616.5} shows the RACS-low (top panel) and -mid (bottom panel) emission centered in \object{BS~J075339$-$2616.5}. In RACS-low, the radio emission appears to be bright and extended, while in RACS-mid, the radio emission is rather compact and weak. The radio source in RACS-low is contained in the RACS Stokes I source catalog \citep{RACSlow2021}, which identifier is \object{RACS\_0741$-$25A\_1914}. The integrated flux is 6.8$\pm$2.1~mJy, and the root mean square (RMS) is $\sim$0.3~mJy~beam$^{-1}$. We computed the RACS-mid flux using a Gaussian function, and obtained $2.0\pm0.5$~mJy. We also estimated the RMS and found $\sim$0.2~mJy~beam$^{-1}$. Searching in SIMBAD at the position of \object{BS~J075339$-$2616.5} we only found the IRAS source \object{IRAS~07515$-$2608}. Therefore, there appear to be no known astrophysical sources in the region associated with this radio source. However, the peak position and the morphology of the radio source do not follow the ones of the IR emission, and we cannot confirm that it is associated with the \object{BS~J075339$-$2616.5} bow shock candidate.

\subsubsection{BS/BU-GR51 (BD $-$08 4623)} \label{Sec:App_Radio_GR51}

\begin{figure}[hbt!]
\centerline{\includegraphics[width=0.82\hsize]{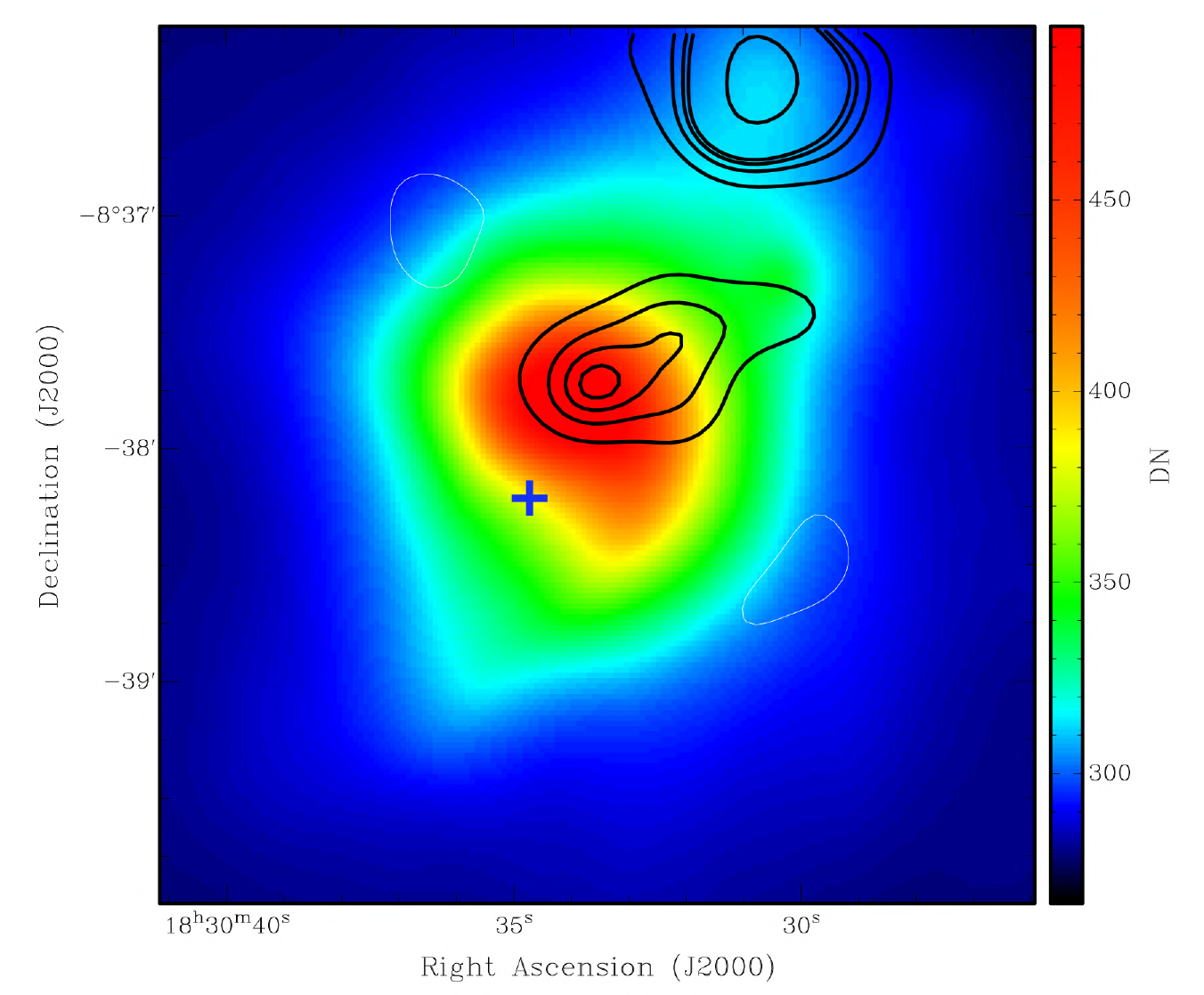}}
\centerline{\includegraphics[width=0.82\hsize]{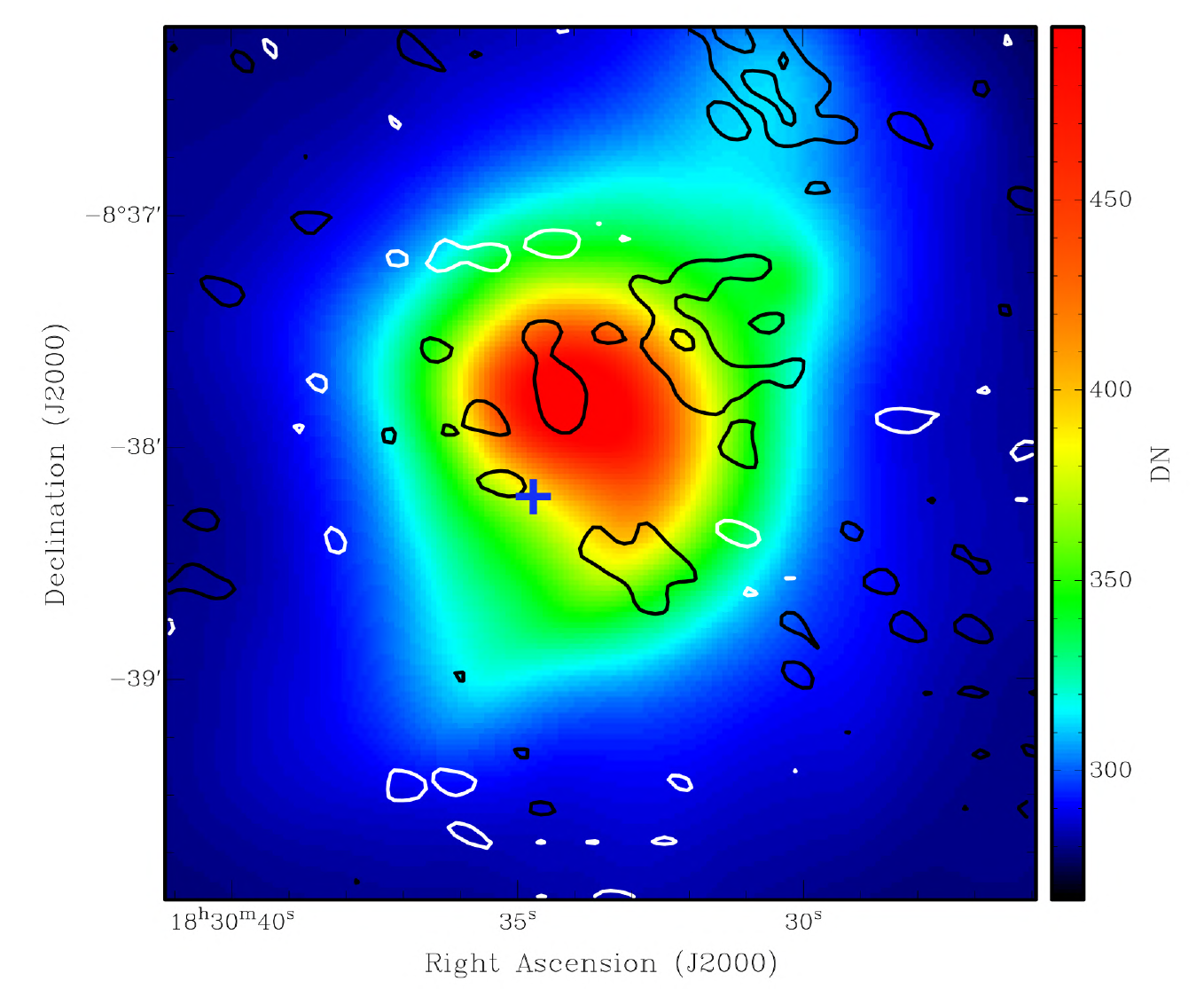}}
\caption{WISE W4 image of the bow shock candidate BS/BU-GR51. The color scale is WISE flux measured in profile-fit photometry, in units of Digital Numbers (DN; see \url{https://wise2.ipac.caltech.edu/docs/release/allsky/expsup/sec1_4c.html\#photcal}). {\sl Top}: RACS-low emission contours onto W4, with levels at $-$2.1 (in white), 2.1, 3, 3.5, 3.8 and 6~mJy~beam$^{-1}$ (in black); synthesized beam of $25'' \times 25''$. {\sl Bottom}: RACS-mid emission contours onto W4, with levels at $-$0.5 (in white), 0.5 and 1~mJy~beam$^{-1}$ (in black); synthesized beam of $10.9'' \times 8.2''$. The cross indicates the position of the runaway star.}
\label{Fig:BSGR51_RACS}
\end{figure}

Although there is no radio emission in VLASS, there is in NVSS and RACS images at the position of BS/BU-GR51. There are actually two sources, one at the position of the bow shock and another brighter one to the North; see the RACS-low and -mid images presented in top and bottom panels of Fig.~\ref{Fig:BSGR51_RACS}, respectively. The northern source appears in \cite{RACSlow2021} catalog, \object{RACS\_1824-06A\_1653}, and does not seem to be related to our bow shock. In RACS-low, there is radio emission close to the center of the bow shock. However, the morphology does not seem to follow that of the bow shock in W4. Therefore, we leave this source as a likely radio counterpart, but we cannot claim for a clear association. In any case, we measured the radio flux density and found $5.1\pm0.1$~mJy.
 In RACS-mid, there are just some peaks of radio emission at the position of the bow shock, so a possible radio counterpart to the bow shock is doubtful. The flux density measured in RACS-mid over the 3$\sigma$ (RMS) contour is $0.9\pm0.1$~mJy (RMS = 0.3~mJy~beam$^{-1}$). The fact that the emission decreases with increasing frequency could indicate a possible nonthermal origin of the radio emission. 

\subsubsection{BS-GR83 (HD 155775)} \label{Sec:App_Radio_GR83}

\begin{figure}[hbt!]
\centerline{\includegraphics[width=0.9\hsize]{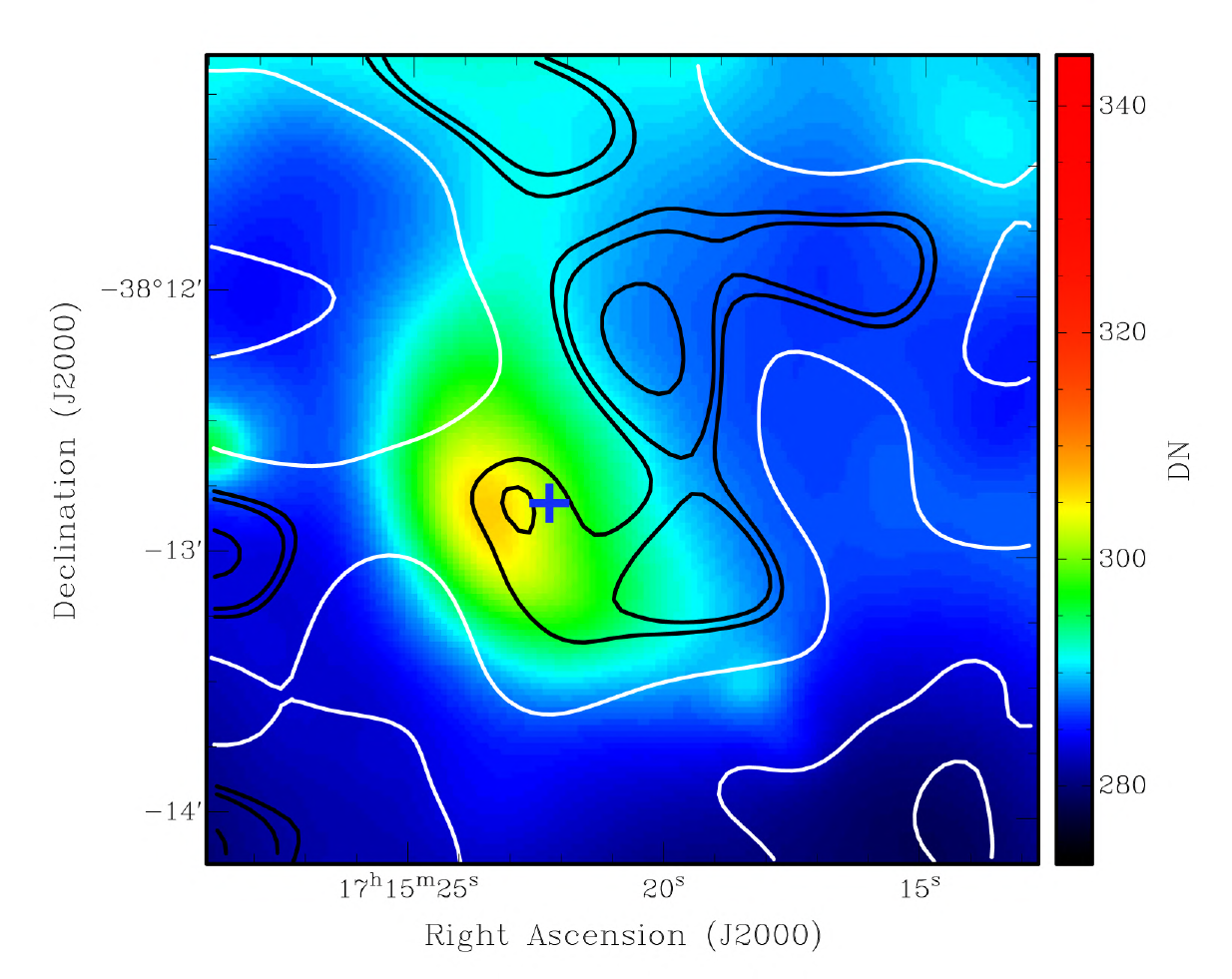}}
\centerline{\includegraphics[width=0.9\hsize]{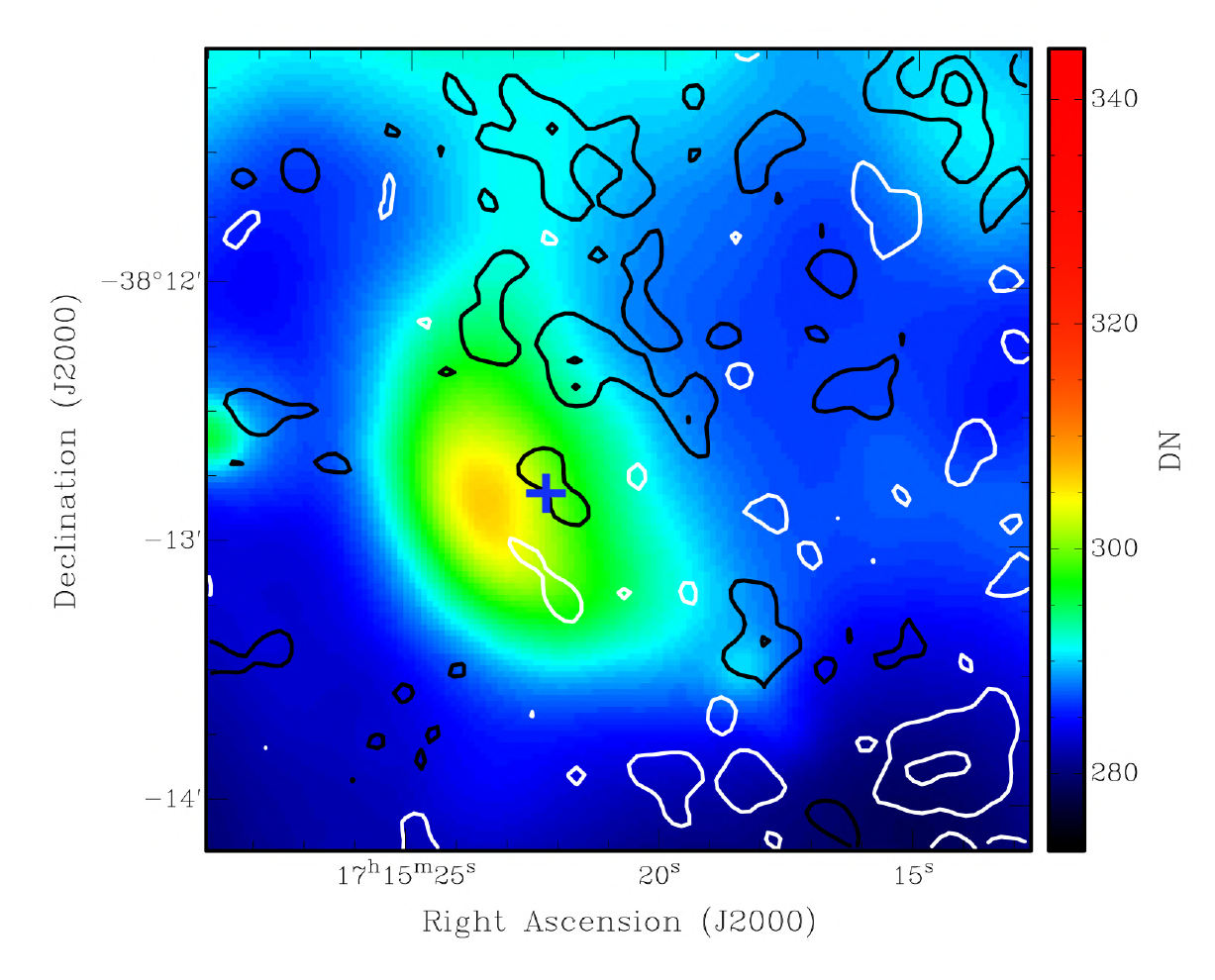}}
\caption{WISE W4 image of the bow shock candidate BS-GR83. The color scale is WISE flux measured in profile-fit photometry, in units of Digital Numbers (DN; see \url{https://wise2.ipac.caltech.edu/docs/release/allsky/expsup/sec1_4c.html\#photcal}). {\sl Top}: RACS-low emission contours onto W4, with levels at $-$1.5 (in white), 1.5, 2.1 and 4~mJy~beam$^{-1}$ (in black); synthesized beam of $25'' \times 25''$. {\sl Bottom}: RACS-mid emission contours onto W4, with levels at $-$0.5 (in white), 0.5 and 1~mJy~beam$^{-1}$ (in black); synthesized beam of $9.7'' \times 8.5''$.}
\label{Fig:BSGR83_RACS}
\end{figure}

For BS-GR83, we did not find images from VLASS. In NVSS, there is no relevant emission superimposed to the bow shock. However, in RACS-low (see Fig.~\ref{Fig:BSGR83_RACS} top panel), there is weak radio emission at the position of the bow shock. In RACS-mid (see Fig.~\ref{Fig:BSGR83_RACS} bottom panel), there are only two very weak peaks. The measured RACS-low flux density is $2.5\pm0.6$~mJy; the RMS of RACS-mid image is 2~mJy~beam$^{-1}$.
We note that both radio maps are quite noisy. Therefore, we cannot claim a radio detection in this case, but there are some radio hints. As in BS-GR75, deeper radio observations could reveal radio emission, and, in this case, it may in fact be nonthermal (see Sect.~\ref{Sec:App_Emission_mechanisms}).

\subsection{Testing radio emission mechanisms for bow shocks} \label{Sec:App_Emission_mechanisms}

Following \cite{VandenEijnden2022b}, we can consider the nonthermal (synchrotron) and the thermal (free-free) scenarios to estimate the expected peak radio flux densities for the non-detected bow shocks. For this, we need the observables from radio images in conjunction with geometrical IR measurements ($R$, $w$), stellar parameters ($\dot{M}$, $\varv_\infty$), distances, and the estimated ISM density ${n_{\mathrm{ISM}}^\mathrm{3D}}$\footnote{We note that in this section we only used the 3D ISM density given that it is the one used by \cite{VandenEijnden2022b}.}. In our work, we already obtained all the necessary ingredients of the IR characterization for the bow shock candidates (see Tables~\ref{Tab:NewBowShocks}~and~\ref{Tab:KnownBowShocks}). 

To compute the radio observables, we used RACS-low DR1 images corresponding to a frequency of 887.5 MHz (34 cm). We accessed the images through the CSIRO Data Access Portal\footnote{\url{https://data.csiro.au/domain/casdaObservation}}. There are two types of RACS images: raw data and final data products. The raw data have reduced sensitivity towards the edge of the field, but provide higher resolution for any individual tile. The final data products are obtained by convolving to a common resolution of 25\arcsec. \cite{VandenEijnden2022b} worked with the raw data products because the final ones were not available at the time of writing. We worked with the two for comparison, but here we only present the raw data to be coherent with that work. The differences we found between the two data products are: elliptic and smaller beam sizes in raw images, and larger RMS values in final images. The RMS is always larger in final data products: depending on the field we found relative differences from 6 to 86\%, with a mean value of 40\%. In Table~\ref{Tab:RadioObs}, we present the radio observables needed from RACS-low (raw) images: the RMS measured in a circular region of 2\arcmin\ centered at the bow shock position, the full image RMS, the beam major and beam minor axes, and the field name of the RACS image used.

\renewcommand{\arraystretch}{1.1}
\begin{table}
\centering
\tiny
\caption{Radio observables obtained from RACS-low images for new and known bow shock candidates.}
\label{Tab:RadioObs}
{\begin{tabular}{l@{~~~}l@{~~}c@{~~}c@{~~}c@{~~}c@{~~}c@{~~}c@{~}}
\hline\hline \vspace{-2mm}\\
id & Runaway Star & RMS  & Image RMS & BMAJ & BMIN & RACS field
\\
   &          & ($\mu$Jy) & ($\mu$Jy) &  $\left(\arcsec\right)$ & $\left(\arcsec\right)$  & 
\\
\hline \vspace{-2mm}\\
BS-GR9  & HD 41997         & 422  &  272  & 20.0  & 13.6 & 0606$+$18A\\
BS-GR35 & HD 64568         & 461  &  268  & 21.6  & 15.6 & 0741$-$25A\\
BS-GR71 & HD 89137         & 252  &  176  & 14.6  & 13.4 & 1027$-$50A\\
BS-GR75 & HDE 338 916      & 413  &  261  & 22.7 & 13.0  & 1955$+$25A\\
BS-GR93 & HD 116282        & 286  &  317  & 20.1 & 13.0  & 1314$-$62A\\
\hline
BS-GR40  & HD 46573        & 737   & 307  & 23.7 & 16.5  & 0624$+$00A\\
BS-GR70  & CPD $-$26 2716  & 424   & 268 & 21.6  & 15.6  & 0741$-$25A\\
BS-GR83  & HD 155775       & 911   & 376 & 13.6  & 12.8  & 1721$-$37A\\
\hline
\end{tabular}}
\tablefoot{The RMS is measured in a 2\arcmin\ circle centered at the position of the bow shock. Image RMS is median image RMS over the tile. BMAJ and BMIN are the beam's major and minor axis size for each image, respectively. We only included here the radio non-detected bow shock candidates that contain all the stellar parameters and estimated $n_\text{ISM}$ needed. We did not include here BS-GR51 because it has a clear radio detection, neither BS-GR72 because it is not visible by ASKAP. We included BS-GR83 because it does not have a clear radio detection. The horizontal line separates the new bow shock (top) from the known bow shock candidates (bottom).}
\end{table}

In \cite{VandenEijnden2022b}, a nonthermal scenario is set by assuming either a low magnetic field of $B=10$~$\mu$G or the maximum magnetic field $B_\text{max}$ so that the bow shock can form with a compressible stellar wind (see Eq.~(2) in \citealt{VandenEijnden2022b}, and references therein). Other assumptions by these authors are: a high injection efficiency $\eta_e$ of 10\%, a maximum electron energy $E_\text{max}=10^{12}$~eV, and a radio spectral index\footnote{The flux density at a given frequency $S_\nu$ is related to the spectral index $\alpha$ by $S_\nu\propto\nu^{\alpha}$. Note that \cite{VandenEijnden2022b} use the alternative notation $S_\nu\propto\nu^{-\alpha}$ and thus $\alpha=0.5$} $\alpha=-0.5$. \cite{VandenEijnden2022b} also considered a thermal scenario based on thin free-free emission. The electron number density $n_e$ and temperature $T$ of the electron population within a bow shock are the two primary physical properties that determine the thermal free-free emission from a shock. These parameters determine both the radio luminosity of the system and the surface brightness of H$_\alpha$ line emission. Using observational and literature constraints on these quantities, \cite{VandenEijnden2022a} inferred values on $n_e$ and temperature $T$, which were compatible with the range of $T = 6\times10^3$~K or $T = 1.4\times10^5$~K found for a sample of H$_\alpha$ detected bow shocks in \cite{BrownBomans2005}. This was the temperature range employed by \cite{VandenEijnden2022b} and subsequently by us in this work. For $n_e$, it is expected a shock density enhancement in the bow shock. \cite{VandenEijnden2022b} assumed $n_e=4n_{\text{ISM}}$, based on the 4 factor predicted by Rankine–Hugoniot equations \citep{Landau1959} (for more details on the scenarios considered and/or the assumptions, see the original work by \cite{VandenEijnden2022b}).

Thanks to the data availability from \cite{VandenEijnden2022b}, we can use their published code\footnote{\url{https://github.com/jvandeneijnden/RACSRadioBowshocks}} to easily reproduce their results and integrate ours. To implement Fig.~10 of this work with our data, we only took the bow shock (not bubbles) candidates in Tables~\ref{Tab:NewBowShocks}~and~\ref{Tab:KnownBowShocks} that have estimated values of ${n_{\mathrm{ISM}}^\mathrm{3D}}$, RACS-low images available, and are non-detected bow shocks. They are BS-GR9, BS-GR35, BS-GR71, BS-GR75, BS-GR93; BS-GR40, BS-GR70, and BS-GR83. We excluded BS-GR51 since it has a clear radio detection, although it does not appear to be associated with the bow shock. We did not exclude BS-GR83 because its possible radio detection is not clear.

\begin{figure}
\centerline{\includegraphics[width=0.91\hsize]{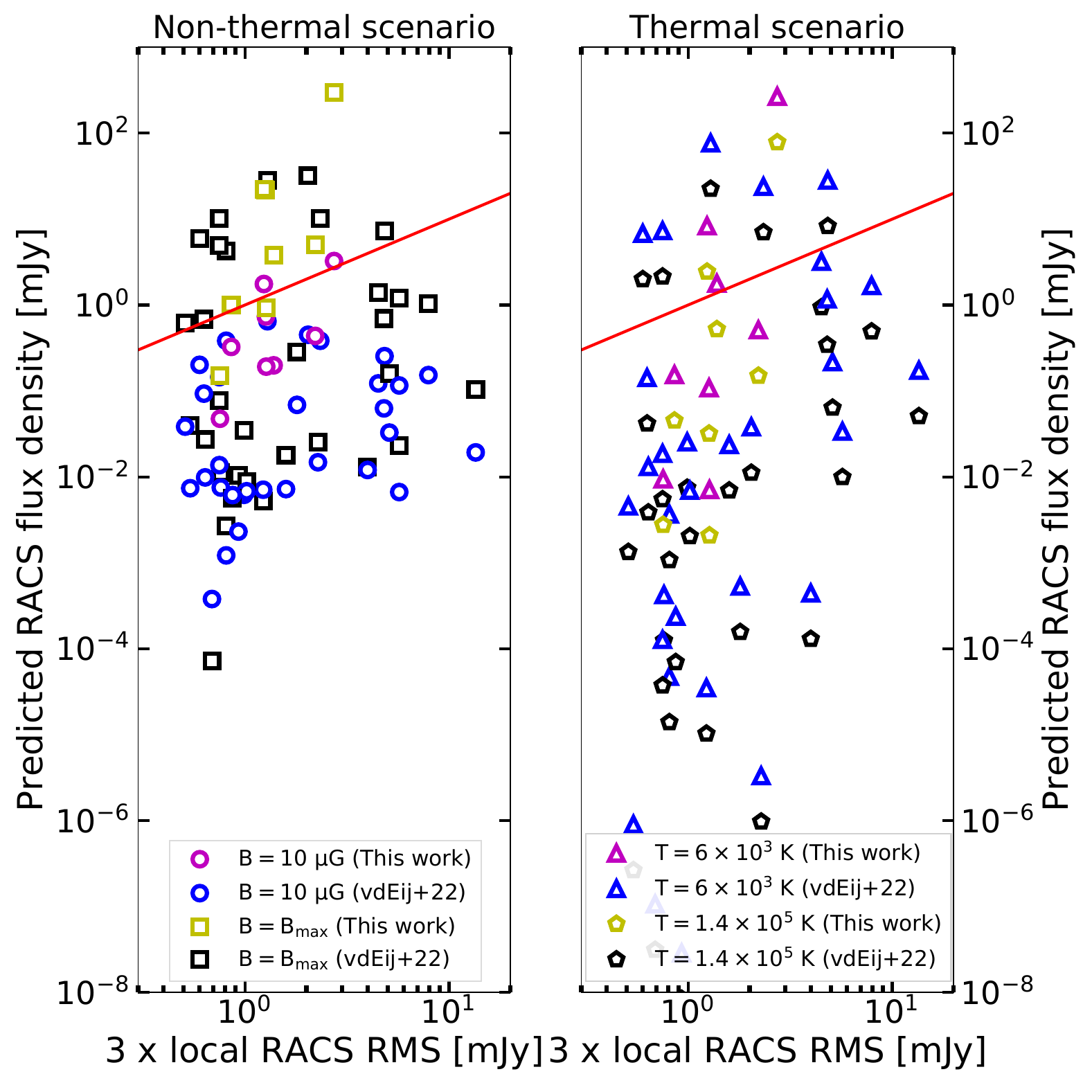}}
\caption{Predicted peak radio flux density as a function of three times the local RMS for the nonthermal (left) and thermal (right) scenarios. This plot contains the original data presented in \cite{VandenEijnden2022b}, and our sources in yellow and magenta colors, with different symbols depending on the scenario conditions. The one-to-one line is shown by the red line; sources above it should, at minimum, have observable emission in a single beam.}
\label{Fig:PredictedRadioFlux}
\end{figure}

The predicted peak radio flux densities for our bow shocks in the nonthermal and thermal scenarios are presented in Fig.~\ref{Fig:PredictedRadioFlux}. They are plotted as a function of three times the local RMS of each bow shock. Left panel corresponds to the nonthermal scenario, while right panel to the thermal one. The different markers show the different conditions in $B$ or $T$ of the scenarios, and also differentiate \cite{VandenEijnden2022b} sources of ours. There are two markers for each source: circles and squares representing the minimum and maximum considered values of $B$, respectively (left panel); and triangles and pentagons for the minimum and maximum considered values $T$, respectively (right panel). Therefore, for the nonthermal scenario, the minimum and maximum predicted peak radio flux density for each source is given by a circle and a square, respectively (except for 3 cases where 
\citealt{VandenEijnden2022b} found $B_\text{max} < 10 $~$\mu$G). In the thermal scenario, the minimum and maximum predicted peak radio flux for each source is given by a pentagon and a triangle, respectively. In both panels the red line denotes where the two flux densities are equal, i.e., sources above that line are expected to be detected in RACS-low images. Thus, the figure gives an idea of how feasible it would be to detect a source in one scenario or the other. For the nonthermal scenario, 6 of our sources should be detectable for the maximum $B$ (two yellow squares are overlapping), and two sources, BS-GR75 and BS-GR83, should be detectable for both magnetic fields. This is the main difference with respect to \cite{VandenEijnden2022b} sources, where they only found detectable sources for the maximum $B$. For the thermal scenario, we find the same two sources detectable for both electron temperatures, and BS-GR35 barely for the minimum one. The non-detection of BS-GR75 and BS-GR83 could be attributed to i) the significant fine tuning required for $\dot{M}$, $\varv_\infty$, and $n_{\text{ISM}}$, to have a magnetic field close to $B_\text{max}$; ii) the lower limit of $B=10$~$\mu$G is susceptible to small variations (for $B=8$~$\mu$G the sources would not be detectable); iii) the possible overestimation of $n_{\text{ISM}}$ that would leave the sources undetectable in the thermal scenario. Explanation iii) could play an important role when working with projected distances (as it is our case here), because of the inverse relation between $R$ and $n_{\text{ISM}}$.

The non-detection of BS-GR83 could be questionable, since it exhibits doubtful emission at the bow shock position (see Sect.~\ref{Sec:App_Radio_GR83}). A possible explanation for its unclear detection could be its noisy radio field (see Fig.~\ref{Fig:BSGR83_RACS}). In case the radio source is confirmed, it seems to fit better the nonthermal scenario, since the predicted flux density in the thermal scenario is strongly influenced by a very high ${n_{\mathrm{ISM}}^\mathrm{3D}}$ value. In the case of the non-detection of BS-GR75, the RACS-low image shows a strong radio source near its position. This bright source could hamper the detection of our source. Contrary to BS-GR83, in the case of BS-GR75 it is more difficult to choose one scenario or the other, since its ${n_{\mathrm{ISM}}^\mathrm{3D}}$ does not seem to be so clearly overestimated. More detailed radio observations on these two sources could probably lead to radio detections.

Lastly, we comment on BS-GR51. This source has a radio detection, and there are hints of nonthermal emission (see Sect.~\ref{Sec:App_Radio_GR51}). Moreover, its $R$ is also small. The radio flux density obtained for this source is 5.1~mJy. Therefore, following the above work but in the inverse sense, and assuming a nonthermal scenario, we can estimate as 4.3~$\mu$G the magnetic field required to explain the BS-GR51 flux density. This is a reasonable magnetic field of the order of the typical value of 10~$\mu$G \citep{Gabici2007,Martinez2023}, and therefore the nonthermal scenario could explain the observed flux density. However, we note here again that it is not clear if this radio source is associated with the bow shock.

\subsection{Summary of radio emission searches}
\label{Sec:App_Summary_radio}

We searched in different archives for observational evidence of radio emission coming from our sources. In particular, we used data from NVSS, VLASS and RACS radio surveys. For \object{BS~J075339$-$2616.5}, and BS-GR51, we found radio emission coincident to the position of the bow shock. However, the morphology of the radio emission differs to some extent from that of the WISE bow shock. Therefore, we cannot claim that they have actual radio-emitting counterparts. For BS-GR75 and BS-GR83, there are some hints of radio emission, but their noisy fields make it difficult to assess whether they have real radio detections. BS-GR72 presented radio emission coincident to the bow shock position. However, it is associated with the \ion{H}{II} region S206, and it is indeed thermal radio emission \citep{Omar2002}. Finally, for BS-GR9 we found radio emission but at $\sim$~4\arcmin\ south from it. This radio emission is also associated with a \ion{H}{II} region, in this case, the Lower's Nebula Sh2-261 \citep{Wang2023}. We also investigated a thermal (free-free) and nonthermal (synchrotron) scenario for the origin of the radio emission. We computed a predicted radio flux density of our non-detected bow shocks. We found that two sources, BS-GR75 and BS-GR83, should be detectable in both scenarios and under all conditions considered. Their non-detection could be explained by the difficulty to reach $B_\mathrm{max}$, small variations in the lower limit of $B$, and a possible overestimation of $n_{\mathrm{ISM}}^{\mathrm{3D}}$. However, taking into account the geometrical characterization of these sources (small $R$), and the observational radio results, they could still be compatible with nonthermal radio-emitting sources.

All these leave us with a clear conclusion: more sensitive observations are needed to detect bow shocks at radio wavelengths, and clarify the physical mechanisms that could give rise to the radio emission. Deep radio observations with more sensitive instruments as VLA could uncover a new population of as yet non-detected radio bow shocks, and shed light on their radio and HE emission connection.

\onecolumn
\section{Miscellaneous IR structures} \label{Sec:App_Miscellaneous}

\begin{figure*}[hbt!]
\centerline{\includegraphics[width=\textwidth]{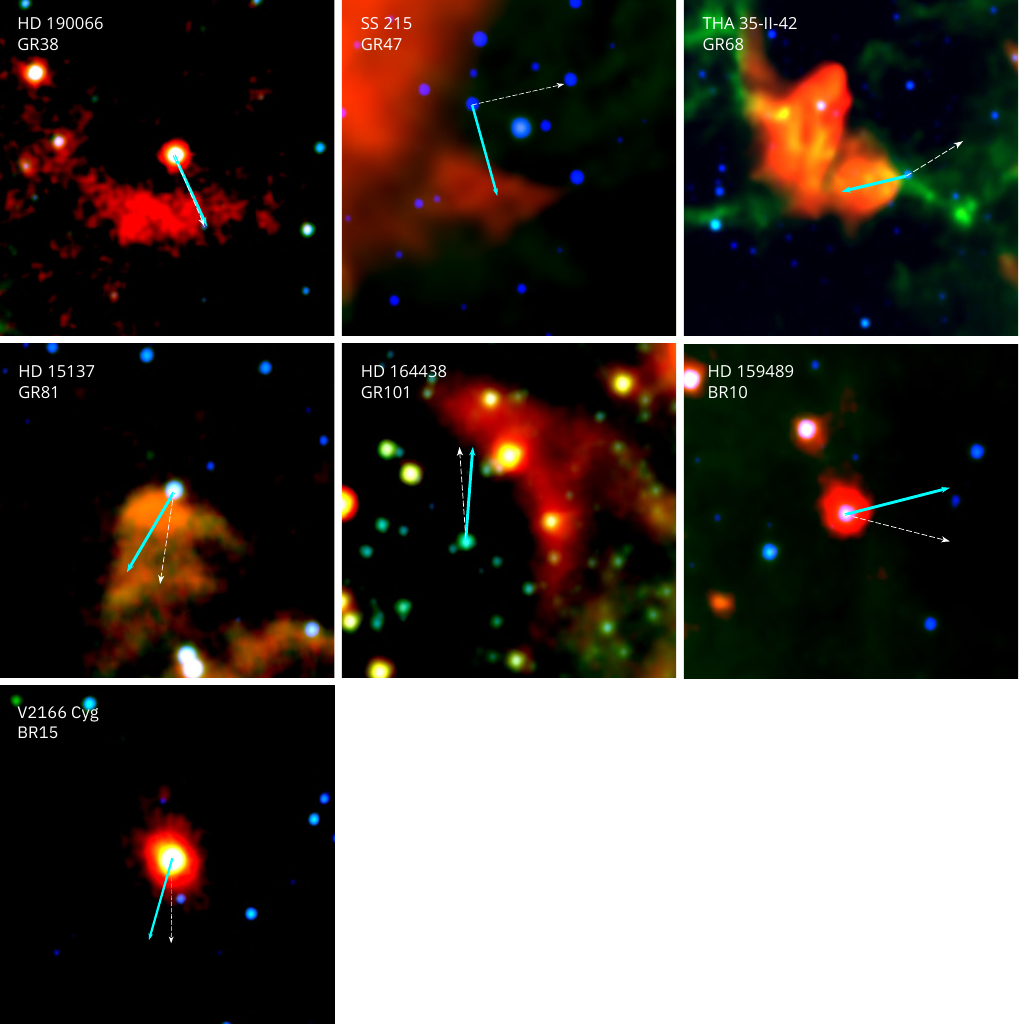}}
\caption{WISE RGB images in W4+W3+W2 for the miscellaneous IR structures presented in Table~\ref{Tab:Miscellaneous} (except for the point-like sources) in equatorial coordinates with North up and East to the left. Dashed white arrows indicate the directions of the original proper motions from \textit{Gaia}~DR3, and solid cyan arrows indicate the directions of the proper motions corrected for the ISM motion caused by Galactic rotation. Each field has a different size, but the arrows are fixed to 2\arcmin. Each panel contains the name of the star as provided in the GOSC or the BeSS catalogs, and the runaway star id.}
\label{Fig:RGBs_Misc_App}
\end{figure*}

The miscellaneous IR structures found in W4 around some runaway stars were introduced in Sect.~\ref{Sec:Miscellaneous}. In this appendix we include the WISE RGB images in W4+W3+W2 of the sources presented in Table~\ref{Tab:Miscellaneous}, except for the point-like sources. The images are presented in Fig.~\ref{Fig:RGBs_Misc_App}. Some of the sources present complex IR fields (GR47, GR68), doubtful IR structures (GR38, GR101), some are mini-bubbles (BR10, BR15), and another presents a kind of bubble followed by a trail (GR81). Detailed multiwavelength studies on these sources could unveil the nature of the IR emission.\\

On the other hand, from the 62 miscellaneous sources we found, 55 presented nearly point-source IR emission. This emission could be associated with an IR excess from the circumstellar disk, especially in the Be-type stars \citep{Rivinius2013}.

\end{appendix}

\end{document}